\documentclass[11pt,draftcls, onecolumn]{IEEEtran}

\makeatletter
\def\ps@headings{%
\def\@oddhead{\mbox{}\scriptsize\rightmark \hfil \thepage}%
\def\@evenhead{\scriptsize\thepage \hfil \leftmark\mbox{}}%
\def\@oddfoot{}%
\def\@evenfoot{}}
\makeatother
\usepackage{multirow}
\usepackage{amsmath,amsthm,amssymb,amsfonts}
\usepackage{tabularx}
\usepackage{graphicx}
\usepackage{subfigure}
\usepackage{hyperref}
\usepackage{url}
\usepackage{balance}
\usepackage{hyperref}
\usepackage{cases}
\usepackage{bbm}
\usepackage{bm}
\usepackage[square, comma, sort&compress, numbers]{natbib}

\usepackage{algorithmic}
\usepackage{algorithm}

\theoremstyle{plain} \newtheorem{theorem}{Theorem}
\theoremstyle{plain} 
\theoremstyle{plain} \newtheorem{lemma}{Lemma}
\theoremstyle{plain} \newtheorem{remark}{Remark}
\theoremstyle{plain} \newtheorem{problem}{Problem}

\begin{document}

%\title{Spatiotemporal-aware Principal Component Pursuit for Data Recovery in Wireless Sensor Networks}
\title{An {\em LS-Decomposition} Approach for Robust Data Recovery in Wireless Sensor Networks}

%1570001151

%\author{\IEEEauthorblockN{
%Xiao-Yang Liu\IEEEauthorrefmark{1}\IEEEauthorrefmark{2},
%Linghe Kong\IEEEauthorrefmark{1},
%Yanmin Zhu\IEEEauthorrefmark{1},
%Yu Gu\IEEEauthorrefmark{3},
%Cai Fu\IEEEauthorrefmark{4},
%Min-You Wu\IEEEauthorrefmark{1}\\
%\IEEEauthorblockA{
%\IEEEauthorrefmark{1}Department of Computer Science and Engineering, Shanghai Jiao Tong University, China\\
%\IEEEauthorrefmark{2}Department of Electrical Engineering, Columbia University, USA\\
%\IEEEauthorrefmark{3}Singapore University of Technology and Design Advanced Digital Sciences Center, Singapore\\
%\IEEEauthorrefmark{4}Department of Computer Science and Technology, Huazhong University of Science and Technology, China\\
%Email: {\IEEEauthorrefmark{1}\{yanglet,linghe.kong,yzhu,mwu\}@sjtu.edu.cn,\IEEEauthorrefmark{2}xl2427@columbia.edu,\\
%\IEEEauthorrefmark{3}jasongu@sutd.edu.sg, \IEEEauthorrefmark{4}fucai@hust.edu.cn
%}
%}
%}}

\author{Xiao-Yang Liu, Xiaodong Wang, {\em Fellow, IEEE}, Linghe Kong, Meikang Qiu, \\and Min-You Wu, {\em Senior Member, IEEE}
\IEEEcompsocitemizethanks{\IEEEcompsocthanksitem  X.~Liu and X.~Wang are with the Department of Electrical Engineering, Columbia University.
\IEEEcompsocthanksitem  X.~Liu and L.~Kong are with the Department of Computer Science and Engineering, Shanghai Jiao Tong University.
\IEEEcompsocthanksitem  M.~Qiu is with the Department of Computer Science, Pace University.}
\thanks{}}

% make the title area
\maketitle

\begin{abstract}

    Wireless sensor networks are widely adopted in military, civilian and commercial applications, which fuels an exponential explosion of sensory data. However, a major challenge to deploy effective sensing systems is the presence of {\em massive missing entries, measurement noise, and anomaly readings}. Existing works assume that sensory data matrices have low-rank structures. This does not hold in reality due to anomaly readings, causing serious performance degradation. In this paper, we introduce an {\em LS-Decomposition} approach for robust sensory data recovery, which decomposes a corrupted data matrix as the superposition of a low-rank matrix and a sparse anomaly matrix. First, we prove that LS-Decomposition solves a convex program with bounded approximation error. Second, using data sets from the IntelLab, GreenOrbs, and NBDC-CTD projects, we find that sensory data matrices contain anomaly readings. Third, we propose an accelerated proximal gradient algorithm and prove that it approximates the optimal solution with convergence rate $O(1/k^2)$ ($k$ is the number of iterations). Evaluations on real data sets show that our scheme achieves recovery error $\leq 5\%$ for sampling rate $\geq 50\%$ and almost exact recovery for sampling rate $\geq 60\%$, while state-of-the-art methods have error $10\% \sim 15\%$ at sampling rate $90\%$.

\end{abstract}


\section{Introduction}

  Wireless sensor networks (WSNs) \cite{Ian2002CN} are constantly generating an enormous amount of rich and diverse information. WSNs are widely adopted in military, civilian and commercial applications, such as intrusion detection in battlefields \cite{Wang2013TPDS}, search and rescue systems \cite{Yugu2014INFOCOM}, infrastructure monitoring \cite{Liu2014INFOCOM}, environment monitoring  etc. The increasing number of big data sources have fueled an exponential explosion of sensory data \cite{Baraniuk2011Sience}.
  Utilizing such a huge amount of sensory data for information substraction and interpretation, we are able to quantitatively understand physical phenomena and perform actively control over cyber-physical systems (CPS).

  However, a major challenge to deploy effective sensor systems is the presence of {\em massive missing entries, measurement noise, and anomaly readings}.
  Missing entries will detriment the usability and reliability of the sensory data sets, while measurement noise and anomaly readings will cause erroneous conclusions and decisions. The data loss rate in real-world projects can be as high as $5\% \sim 95\%$, as shown in Fig. \ref{fig:loss_rate} Section V, due to unreliable wireless communications such as poor link quality and packet collision. Moreover, measurement noise and anomaly readings are ubiquitous in real-world projects mainly due to: (1) commodity sensors have low accuracy; (2) some sensor nodes have malfunctions; and (3) the dynamic surroundings may cause disturbances. Therefore, it is necessary to distinguish environmental data from anomalies readings and measurement noise in a robust and accurate way.

\begin{figure}[t]
  \centering
  \includegraphics[width=0.6\textwidth]{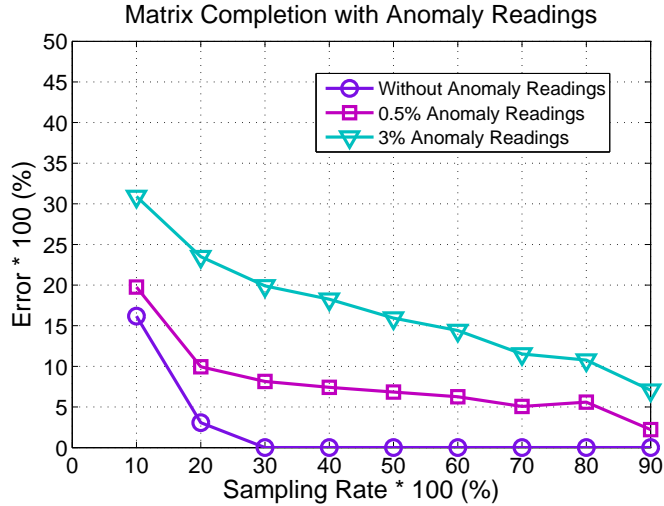}
  \vspace{-8pt}
  \caption{Performance of matrix completion: exactly low-rank matrix, compared with matrices with $0.5\%$ and $3\%$ anomalies.}\label{fig:ls_motivation}
  \vspace{-14pt}
  \end{figure}

  Network data are naturally represented as a matrix in which rows denote nodes and columns denote time slots. Therefore, many recent works apply matrix completion techniques to perform data recovery \cite{Kong2013INFOCOM,Chen2013GlobeCom,Yang2012,Liu2012SECON} and data collection \cite{Cheng2013TWC,Xin2014ICDCS} in WSNs. Compression is performed in the data collection phase to reduce communication burden, while matrix completion is applied at the sink node to recover the raw sensory data. However, the standard matrix completion \cite{Candy2009,Tao2010,TFOCS2011} experiences serious performance degradation with the presence of a small portion of anomalies. Fig. \ref{fig:ls_motivation} shows a simple example, a $100 \times 100$ matrix with rank $r=5$ taking values in $[0,100]$, can be exactly recovered using $30\%$ uniformly random selected samples \cite{TFOCS2011}. However, if we add anomalies with value $=100$, two difference ratios $0.5\%$ and $3\%$, the recovery errors ($\ell_2$-norm) increase significantly.

\begin{figure}[t]
  \centering
  \includegraphics[width=0.6\textwidth]{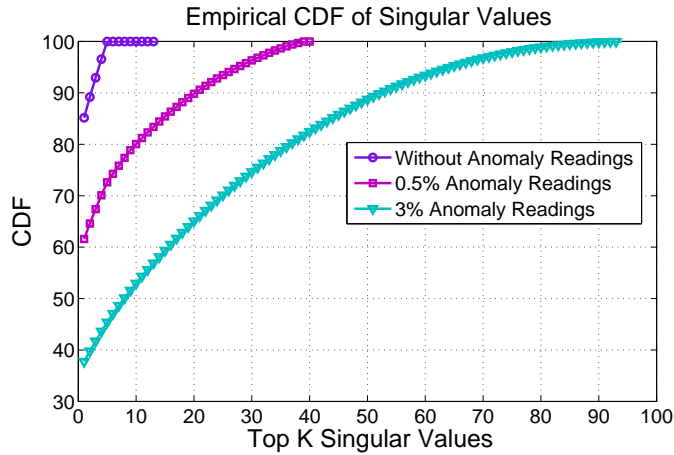}
  \vspace{-8pt}
  \caption{Empirical CDF of energy captured by the top $K$ singular values: exactly low-rank matrix, compared with matrix with $0.5\%$ and $3\%$ anomalies.}\label{fig:ls_motivation_svd}
  \vspace{-14pt}
  \end{figure}

  The performance degradation comes from the fact that: even a small portion of anomaly readings can break down the low-rank structure. As shown in Fig. \ref{fig:ls_motivation_svd}, adding $0.5\%$ anomalies into the above rank $r=5$ matrix, the energy captured in the top $10$ singular values is $80\%$, while it is only $53\%$ when $3\%$ anomalies are added. In Section V, using data sets collected from the IntelLab \cite{Intel}, GreenOrbs \cite{liu2011does}, and NBDC-CTD \cite{NBDC} projects, we observe that anomaly readings are common and ubiquitous. Therefore, we find that existing works hastily assume low-rank structures.

  In this paper, we introduce an {\em LS-Decomposition} approach for sensory data recovery, which can deal with massive data loss and is robust to both measurement noise and anomaly readings. By modeling measurement noise as small entry-wise noise, and anomaly reading as gross sparse error, LS-Decomposition decomposes a sensory data matrix into a low-rank matrix and a sparse anomaly matrix. Firstly, we present observations on real data sets from the IntelLab \cite{Intel}, GreenOrbs \cite{liu2011does}, and NBDC-CTD \cite{NBDC} projects, showing that (1) the original data matrices have massive missing entries and (2) each data matrix is the superposition of a low-rank matrix and a sparse anomaly matrix. Secondly, we formulate the robust data recovery problem as an optimization problem, coined as {\em LS-Decomposition}, and prove that solving a convex program achieves bounded approximation. Thirdly, we propose an accelerated proximal gradient algorithm for LS-Decomposition, theoretical results guarantee that our algorithm approximates the optimal solution with convergence rate $O(1/k^2)$ ($k$ is the number of iterations). Finally, evaluations on real data sets show that our scheme achieves recovery error $\leq 5\%$ for sampling rate $\geq 50\%$ and almost exact recovery for sampling rate $\geq 60\%$, while state-of-the-art methods have error $10\% \sim 15\%$ at sampling rate $90\%$.

  The reminder of the paper is organized as follows. Section II discusses related works. System models and problem formulation are given in Section III. We present conditions and theoretic guarantees for optimal LS-Decomposition in Section IV. Observations are presented in Section V. Our algorithm is in Section VI, while Section VII describes performance evaluation. We conclude in Section VIII. The appendix provides detailed proofs.

\section{Related Work}

 We first describe the applications of the standard matrix completion in wireless sensor networks. Then, we discuss several variants of matrix completion.

 Data loss is revealed to be ubiquitous and unavoidable in sensor networks \cite{Kong2013INFOCOM}. In order to improve the reliability and useability of decisions draw from such incomplete sensory data, a data recovery process is needed. Since the raw sensory data are redundant \cite{Correlation2004}, it is able to estimate the original environment data from partial observations. In \cite{Kong2013INFOCOM}, authors show that data matrices of temperature, humidity and light are low-rank and have high spatiotemporal correlations, while similar empirical observations are also presented in \cite{Xin2014ICDCS}. To recover effectively, \cite{Kong2013INFOCOM,Liu2012SECON} apply data preprocess on the raw data sets to exclude anomaly readings, and then matrix completion to perform data recovery. However, there are no good criteria for identifying anomaly readings while \cite{Kong2013INFOCOM,Liu2012SECON} are based on researchers' experience. Furthermore, the inter-correlation among multiple attributes are exploited to improve the recovery accuracy \cite{Chen2013GlobeCom}. Matrix completion is also suitable for seismic data recovery \cite{Yang2012}.

 Besides data recovery, matrix completion is also applied to improve data collection \cite{Cheng2013TWC,Xin2014ICDCS} in WSNs. Raw-data collection is rather inefficient \cite{Fazel2011JSAC}, since WSNs are typically composed of hundreds to thousands of sensor nodes generating tremendous amount of sensory readings. As the packet loss problem and the hot spot problem surface, this approach will lead to a large number of retransmissions in real-world situations and node failures as  batteries run out. Therefore, researchers apply matrix completion to reduce global traffic \cite{Cheng2013TWC}. Compression is performed in the data collection phase to reduce communication burden, while matrix completion is applied at the sink node to recover the raw sensory data. Furthermore, the authors of \cite{Xin2014ICDCS} propose an online data gathering scheme for weather data.

 A natural technique to improve the recovery performance of the standard matrix completion is to add a smooth regulation term \cite{Kong2013INFOCOM,Chen2013GlobeCom,Liu2012SECON,Xin2014ICDCS}. This intuition is inspired by the fact that almost all physical conditions are smooth fields, i.e., the physical conditions are continuous without sudden changes. However, all the above schemes are based on a critical assumption that the matrix of interest exhibits a low-rank structure, which does not hold in reality. As shown in Fig. \ref{fig:ls_motivation_svd}, \ref{fig:loss_rate} and \ref{fig:svd}, real-world data matrices violate such assumption due to the existence of a small portion of anomaly readings. Therefore, the standard matrix completion \cite{Candy2009,Tao2010} and the smoothed counterpart experience serious performance degradation in practical scenarios.

 Noticing the ubiquitousness of anomalies in practical scenarios, researchers propose to decompose the data matrix into a low-rank matrix and a sparse anomaly matrix \cite{Wright2011,Wright2009,Wright2010Stable} and prove its universal applicability. The authors prove that it is possible to exactly separate those two components. These works inspire us in proving that our problem has bounded approximation, however, our work essentially differs from theirs in terms of modells and goals: (1) no measurement noise is allowed in their model. They consider an {\em exactly} low-rank matrix, while we deal with an {\em approximately} low-rank matrix which is the case for sensory data matrices; (2) full observations of all entries are required. They target at the possibility of {\em separating} two matrices while we aim to {\em recover} the low-rank matrix.

 %Upon the submission, we notice that a recent work \cite{Qiu2014} proposed a similar approach for robust data recovery. However, there are two major differences: (1) we target specifically at sensor network data recovery while \cite{Qiu2014} deals with network traffic and RSS (received signal strength) matrices; (2) \cite{Qiu2014} adopts an alternating direction method which is a purely heuristic approach, while we provide theoretical guarantees for bounded approximation error.

\section{System Model and Problem Statement}

\subsection{Notations}

  We go over the notations and preliminaries.
  Throughout the paper, $N$ denotes the number of nodes in the wireless sensor network, $T$ denotes the number of slots in the sensing period of interest, $[N]$ denotes the set $\{1,2,...,N\}$, $R \in \mathbb{R}^{N \times T}$ denotes the data matrix of interest, $L \in \mathbb{R}^{N \times T}$ denotes the low-rank matrix of interest, and $S \in \mathbb{R}^{N \times T}$ denotes the anomaly matrix of interest. Let  $\Omega \subset [N] \times [T]$ denote the support set of S (i.e., the index set of the nonzero entries), $\mathcal{O} \subset [N] \times [T]$ denote the index set of the observed entries. $R^*$ denotes the transpose of $R$. The operator $\mathcal{P}_{\mathcal{O}}(R)$ projects matrix $R$ onto its index set $\mathcal{O}$, i.e., the $(n,t)$-th entry of $\mathcal{P}_{\mathcal{O}}(R)$ is equal to $R_{nt}$ if $(n,t) \in \mathcal{O}$ and zero otherwise.

  The $\ell_0$-norm is defined as $||S||_0 \doteq |\text{supp}(S)|$, where $\text{supp}(S) \doteq \{(n,t): S_{nt} \neq 0 \}$ denotes the support set of $S$ and $|\text{supp}(S)|$ denotes the cardinality of $\text{supp}(S)$.  The $\ell_1$-norm is defined as $||S||_1 \doteq \sum_{n=1}^{N}\sum_{t=1}^{T} \text{abs}(S_{nt})$ where $\text{abs}(S_{nt})$ denotes the absolute value of $S_{nt}$. The $\ell_2$-norm of vector $A \in \mathcal{R}^{m}$ is defined as $||A||_2 = \sqrt{\sum_{i=1}^{m}A_i^2}$.
  The Frobenius norm of matrix $R$ is defined as $||R||_F \doteq \sqrt{ \sum_{n=1}^{N}\sum_{t=1}^{T} R_{nt}^2}$. Let $L = U \Sigma V^* = \sum_{i=1}^{r} \sigma_i u_i v_i^*$ denote the singular value decomposition (SVD) of $L$, where $r$ is the rank, $\sigma_1 \geq ... \geq \sigma_r > 0$ are the singular values, and $U=[u_1,...,u_r], V=[v_1,...,v_r]$ are the matrices of left- and right-singular vectors, respectively. The nuclear norm of $L$ is defined as $||L||_{*} = \sum_{i=1}^{r} \sigma_i$. Given a matrix pair $X=\langle L,S\rangle$, let $||X||_F \triangleq (||L||_F^2 + ||S||_F^2)^{1/2}$.

  More notations are needed in our proofs as in \cite{Wright2011,Wright2009,Wright2010Stable}. Let $\mathcal{T}$ denote the subspace generated by matrices with the same column space or row space of $L$:
  \begin{equation}
  \mathcal{T}=\{ UQ^* + RV^* | Q,R \in \mathbb{R}^{n \times r} \} \subseteq \mathbb{R}^{n \times n},
  \end{equation}
  and $\mathcal{P}_{\mathcal{T}}$ be the projection operator onto the subspace $\mathcal{T}$. Define the projection operator $\mathcal{P}_{\mathcal{T}} \times \mathcal{P}_{\Omega}: (L,S) \longmapsto (\mathcal{P}_{\mathcal{T}} L, \mathcal{P}_{\Omega}S)$.
  Define the subspace $\Gamma \triangleq \{(Q,Q) | Q \in \mathbb{R}^{n \times n} \}$ and $\Gamma^{\bot} \triangleq \{(Q,-Q) | Q \in \mathbb{R}^{n \times n} \}$, and let $\mathcal{P}_{\Gamma}$ and $\mathcal{P}_{\Gamma^{\bot}}$ denote their respective projection operators.

\subsection{Network Model}

   We consider a wireless sensor network consisting of $N$ sensor nodes and a sink. Sensor nodes are scattered in a target field to monitor physical conditions such as temperature, humidity, illumination, gas concentration, magnetic
   strength, etc. They report sensory readings to the sink periodically over a given time span.

   The monitoring period is evenly divided into $T$ time slots, denoted as $\{0,1,...,t,...,T-1\}$. Each sensor node generates a record in each slot. A record at a sensor node includes the sensor readings, node ID, time stamp, and location (longitude and latitude). Its format is:

   \begin{table}[!htp]\centering
    \vspace{-6pt}
   \begin{tabular}{|c|c|c|c|c|c|}
   \hline
   Record: &sensor readings & node ID & time stamp & location  \\
   \hline
   \end{tabular}
   \vspace{-6pt}
   \end{table}

   Let $R_{i,t}$ denote the sensory reading of the $i$-th node at slot $t$. For each physical condition, the sensor readings generated by all sensor nodes can be represented by a matrix $R \in \mathbb{R}^{N \times T}$, as follows:
   \begin{equation}\label{L} \small
   \setlength{\abovedisplayskip}{4pt}
   R=
   \left[
   \begin{array}{ccccc}
   R_{0,0} & ... & R_{0,t} & ... & R_{0,T-1} \\
   R_{1,0} & ... & R_{1,t} & ... & R_{1,T-1} \\
   ... & ... & ... & ...\\
   R_{N-1,0} & ... & R_{N-1,t} & ... & R_{N-1,T-1}
   \end{array}
   \right]
   \vspace{-4pt}
   \end{equation}
   where rows denote nodes, and columns denote time slots.

\subsection{Measurement Model}

   It is widely believed that sensory data exhibit strong spatio-temporal correlation \cite{Correlation2004,Liu2014TPDS}. Compressive sensing theory \cite{Candy2009,Tao2010} introduces the general notion of ``low-rank" to model this characteristic.
   %In the compressive sensing context, the authors assume that sensory data matrix has low-rank structure \cite{Kong2013INFOCOM}\cite{Xin2014ICDCS}\footnote{The authors preprocess their data sets to exclude anomaly readings. But, there is no good criteria for eliminating anomaly readings.}. However, in real world applications, the sensory readings are often corrupted by measurement noise and anomaly readings (as revealed in Section IV). The low-rank structure of the sensory reading matrix $R$ is therefore broken down.
   In this paper, we assume that the actual sensory data of a physical condition is an {\em approximately} low-rank matrix $L \in \mathbb{R}^{N \times T}$ as in \cite{Kong2013INFOCOM,Chen2013GlobeCom,Yang2012,Liu2012SECON,Cheng2013TWC,Xin2014ICDCS}, which means that most of its energy is captured by its rank-$r$ approximation. The observed sensory reading matrix $R$ was generated by corrupting the entries of $L$ with measurement noise and anomaly readings. We model measurement noise as additive small entry-wise noise, and anomaly readings as additive gross sparse errors, respectively, i.e., $Z \in \mathbb{R}^{N \times T}$ with $||Z||_F \leq \delta$ for some $\delta > 0$, and $S \in \mathbb{R}^{N \times T}$ with $||S||_0 \leq k$. It is reasonable to assume that the number of anomaly readings is relatively small, compared with the size of the sensory data matrix, i.e., $k \ll NT$. Let $\Omega \subset [N] \times [T]$ denote the support set of S, i.e., the index set of the nonzero entries. We have the following measurement model:
   \begin{equation}
   \setlength{\abovedisplayskip}{4pt}
   R=L + S + Z.
   \vspace{-4pt}
   \end{equation}
   %where $L$ is an {\em approximately} low-rank matrix, $S$ is a sparse matrix with most of its entries being zero, and $Z$ is a noise term, say i.i.d. noise on each entry of the matrix.
   %All we assume about $Z_0$ in this paper is that $||Z||_F \leq \delta$ for some $\delta > 0$.

\subsection{Problem Statement}

   Sensor nodes cooperate with each other to transmit packets back to the sink via multi-hop wireless communication. Two major factors lead to massive missing entries: (1) in each hop, poor link quality results in decoding failures at the receiver node; (2) along the multi-hop path, packet collisions are unavoidable since wireless communication utilizes unregulated media access.

   Therefore, the sink receives an incomplete measurement matrix $M$. The data loss rate can be as high as $5\% \sim 95\%$ as shown in Fig. \ref{fig:loss_rate} Section \ref{observation}. Suppose the collected entries are indicated by the set $\mathcal{O} \subset [N] \times [T]$ and $\mathcal{O}$ has size $m$. We assume that the missing entries are randomly distributed, either uniformly random or non-uniform random. Please refer to \cite{Wright2011} for more mathematic details about allowed distributions. The data collection process can be represented as:
   \begin{equation}
   \setlength{\abovedisplayskip}{4pt}
   M=\mathcal{P}_{\mathcal{O}}(R),
   \vspace{-4pt}
   \end{equation}
   where the operator $\mathcal{P}_{\mathcal{O}}(X)$ projects $X$ onto the index set $\mathcal{O}$.

   To estimate the low-rank matrix $L$ from the partial observations $M$, a direct conceptual solution is: to seek a $L$ with the lowest-rank that could have generated the data matrix $R$, subject to the constraints that the gross errors are sparse and the entry-wise errors are small. We formulate the {\em LS-Decomposition} problem as follows:
   \begin{problem}
   {\em (LSD)} Assuming $R=L+S+Z$ and given its partial observation $M=\mathcal{P}_{\mathcal{O}}(R)$ on index set $\mathcal{O}$, where $L,S,Z$ are unknown, but $L$ is known to be low-rank, $S$ is known to be sparse, and $||Z||_F \leq \delta$ for some $\delta > 0$, recover $L$. The Lagrangian formulation is:
   \begin{equation}\label{LS-Decomposition}
   \begin{split}
   &\langle \hat{L}, \hat{S}\rangle  = \min\limits_{\langle L, S\rangle}~~rank(L) + \lambda ||S||_0, \\
   &s.t.~~||\mathcal{P}_{\mathcal{O}}(R-L-S)||_F \leq \delta,
   \end{split}
   \end{equation}
   where we choose $\lambda = 1/\sqrt{n}$, and $\hat{L}, \hat{S}$ are estimates of $L,S$, respectively.
   \end{problem}

\section{Existing Results for LS-Separation}
\label{section_LS_Separation}

   A special case of the LSD problem (\ref{LS-Decomposition}) is when we have full observation of $R$, then (\ref{LS-Decomposition}) is reduced to the following {\em LS-Separation} problem studied in \cite{Wright2011,Wright2009, Wright2010Stable}. In this section, we summarize the conditions for unique solution and existing results.
   \begin{problem}
   {\em (LSS)} Assuming $R=L+S+Z$ and given full observation of $R$ as $\mathcal{O} = [N] \times [T]$, recover the components $L$ and $S$. The Lagrangian formulation is:
   \begin{equation}\label{LS-Separation}
   \begin{split}
   \setlength{\abovedisplayskip}{4pt}
   &\langle \hat{L}, \hat{S}\rangle = \min\limits_{\langle L, S\rangle}~~rank(L) + \lambda ||S||_0, \\
   &s.t.~~||R-L-S||_F \leq \delta.
   \vspace{-4pt}
   \end{split}
   \end{equation}
   \end{problem}

  The LSS problem (\ref{LS-Separation}) is a highly nonconvex optimization problem and no efficient solution is known, since (\ref{LS-Separation}) subsumes both the low-rank matrix completion problem and the $\ell_0$-minimization problem. Both of them are NP-hard and hard to approximate \cite{Meka2008NP}. To obtain a tractable optimization problem, one can relax the LSS problem (\ref{LS-Separation}) by replacing the $\ell_0$-norm with the $\ell_1$-norm and the rank function with the nuclear norm as in \cite{candes2006near,Candy2009,candes2006robust,Wright2009,Wright2011,Wright2010Stable}, yielding the following problem:
  \begin{problem}
  The convex surrogate of the LSS problem (\ref{LS-Separation}) is:
  \begin{equation}\label{LS}
  \begin{split}
   &\langle \hat{L}, \hat{S}\rangle  = \min\limits_{\langle L, S\rangle}~~||L||_{*} + \lambda ||S||_1, \\
   &s.t.~~||R-L-S||_F \leq \delta
  \end{split}
  \end{equation}
  \end{problem}

\subsection{Conditions for Unique Solution}

  Before analyzing the optimality of problem (\ref{LS}), one should answer the following a basic question first: When is LS-Separation possible?
  At first sight, one may think that the objective of problem (\ref{LS-Decomposition}) is impractical. For instance, suppose the matrix $R$ is equal to $e_1e_1^*$ ($e_1$ is a canonical basis vectors, the resulting matrix $e_1e_1^*$ has a one in the top left corner and zeros everywhere else), then since $R$ is both sparse and low-rank, how can we decide whether it is low-rank or sparse? To make problem (\ref{LS-Decomposition}) meaningful, we need to impose that the low-rank component $L$ is not sparse, and the sparse component $S$ is not low-rank.

\subsubsection{The Low-Rank Matrix is Not Sparse}

  Let $L = U \Sigma V^* = \sum_{i=1}^{r} \sigma_i u_i v_i^*$ denote the singular value decomposition (SVD) of $L \in \mathbb{R}^{N \times T}$, where $r$ is the rank, $\sigma_1,...,\sigma_r$ are the singular values, and $U=[u_1,...,u_r], V=[v_1,...,v_r]$ are the matrices of left- and right-singular vectors, respectively.

  The incoherence condition introduced in \cite{Candy2009,Tao2010} asserts that for small values of parameter $\mu$, the singular vectors are reasonably spread out, namely, the low-rank matrix $L$ is not sparse. It states that:
  \begin{equation}\label{LS_condition}
  \max\limits_{i} ~|| U^*e_i||^2 \leq \frac{\mu r}{N}, ~~\max\limits_{i} ~|| V^*e_i||^2 \leq \frac{\mu r}{T}, ~~||UV^*||_{\infty} \leq \sqrt{\frac{\mu r}{NT}},
  \end{equation}
  where $e_i$'s are the canonical basis vectors.

\subsubsection{The Sparse Matrix is Not Low-Rank}

  Another identifiability issue arises if the sparse matrix $S \in \mathbb{R}^{N \times T}$ has low-rank. For instance, all the nonzeros entries of $S$ lie in a column or in a few columns, suppose that the columns of $S$ happens to be the opposite of those of $L$, then it is clear that we would not be able to recover $L$ and $S$ by any method whatsoever since $R=L + S$ would have a column space equal to, or included in that of $L$. To avoid such pathological cases, we assume that the support of sparse component $S$ is selected uniformly at random among all subsets of size $k$ as in \cite{Wright2009,Wright2011,Wright2010Stable}.

\subsection{Existing Main Results}

  The main result of \cite{Wright2010Stable} is that under the above conditions, (\ref{LS}) gives a stable estimate of $L$ and $S$.
%
%  \begin{lemma}
%  \cite{Wright2010Stable} Suppose that $L$ obeys (\ref{LS_condition}) and the support of $S$ is uniformly distributed. Then if $rank(L) \leq \rho_rn\mu^{-1}(\log{n})^{-2}$ and $|\Omega| \leq \rho_s n^2$ with $\rho_r,\rho_s > 0$ being sufficiently small numerical constants. With high probability, the solution $(\hat{L},\hat{S})$ to (\ref{LS}) satisfies
%  \begin{equation}
%  ||\hat{L} - L||_F^2 + ||\hat{S} - S||_F^2 \leq C n^2 \delta^2,
%  \end{equation}
%  where C is a numerical constant, and $n= \max(N,T)$.
%  \end{lemma}

  \begin{lemma}
  \cite{Wright2010Stable} Suppose that $L$ obeys (\ref{LS_condition}) and the support of $S$ is uniformly distributed. Then if $rank(L) \leq \rho_rn\mu^{-1}(\log{n})^{-2}$ and $|\Omega| \leq \rho_s n^2$ with $\rho_r,\rho_s > 0$ being sufficiently small numerical constants. Assume $||\mathcal{P}_{\Omega} \mathcal{P}_{\mathcal{T}} ||\leq 1/2, \lambda \leq 1/2$, let $\hat{X} = \langle \hat{L}, \hat{S} \rangle$ be the solution to (\ref{LS}). Then with high probability, $\hat{X}$ satisfies:
  \begin{equation}
  ||\hat{X} - X ||_F = ||\hat{L}+\hat{S}-L-S||_F \leq (8\sqrt{5}n + \sqrt{2})\delta,
  \end{equation}
  where $n= \max(N,T)$.
  \end{lemma}

  We can see that (\ref{LS}) is simultaneously stable to small entry-wise noise and robust to gross sparse anomalies. Note that the error bound for each entry is proportional to the noise level $\delta$. As for ``with high probability", there is no exact result while \cite{Wright2009,Wright2011,Wright2010Stable} proved that the probability is at least $1- cn^{-10}$ where $c$ is a numerical constant.

\section{Conditions and Performance Guarantees for LS-Decomposition}

\subsection{Problem Analysis}

%  Problem (P\ref{LS-Decomposition}) is a highly nonconvex optimization problem, and no efficient solution is known. Because (P\ref{LS-Decomposition}) subsumes both the low-rank matrix completion problem and the $\ell_0$-minimization problem, both of which are NP-hard and hard to approximate. We relax it to obtain a tractable optimization problem by replacing the $\ell_0$-norm with the $\ell_1$-norm, and the rank function with the nuclear norm $||X||_{*}=\sum_{i} \sigma_{k}(X)$ as in \cite{candes2006near,Candy2009,candes2006robust,Wright2009,Wright2011,Wright2010Stable}, yielding the following problem:

  Following the framework in Section \ref{section_LS_Separation}, we relax problem (\ref{LS-Decomposition}) by replacing the $\ell_0$-norm with the $\ell_1$-norm and the rank function with the nuclear norm. Then, we provide a condition for unique solution and the corresponding theoretical guarantees.

  \begin{problem}
  The convex surrogate of the LSD problem (\ref{LS-Decomposition}) is:
  \begin{equation}\label{LSD}
  \begin{split}
   &\langle \hat{L}, \hat{S}\rangle  = \min\limits_{\langle L, S\rangle}~~||L||_{*} + \lambda ||S||_1, \\
   &s.t.~~||\mathcal{P}_{\mathcal{O}}(R-L-S)||_F \leq \delta
  \end{split}
  \end{equation}
  \end{problem}

\subsection{Partial Observation}

  However, those two conditions in Section \ref{section_LS_Separation} do not suffice to guarantee unique solution for problem (\ref{LS-Decomposition}). For example, all $\{ \langle L, S \rangle, \langle L, S_1 \rangle,...,\langle L, S_n \rangle \}$ are optimal solutions as long as $\mathcal{P}_{\mathcal{O}}(S_1)=... \mathcal{P}_{\mathcal{O}}(S_n)=\mathcal{P}_{\mathcal{O}}(S)$.
  To avoid such predicament, we set the following artificial setting for (\ref{LS-Decomposition}) and (\ref{LS}) throughout the paper:
  \begin{equation}\label{condition}
  || \mathcal{P}_{\mathcal{O}^{\bot}(\hat{S})} ||_F =  || \mathcal{P}_{\mathcal{O}^{\bot}(S_1)} ||_F =...=  || \mathcal{P}_{\mathcal{O}^{\bot}(S)} ||_F = 0.
  \end{equation}
  This is quite reasonable because our aim is to recover the low-rank matrix $L$ and we do not want to recover the anomaly matrix.

  \begin{remark}
    Partial observation of $R$ provides partial recovery of its sparse component $S$, i.e., only the corresponding subset of entries are observable.
  \end{remark}

\subsection{Theoretic Guarantees}

  Under the above conditions, we have the following theoretical results: 1) problem (\ref{LS-Decomposition}) is possible by solving the convex program (\ref{LSD}), and 2) the precise closed form of the approximation bound. The detailed proofs are given in the appendix.
%  \begin{theorem}\label{Main_results}
%  Suppose that $L$ obeys (\ref{LS_condition}) and the support of $S$ is uniformly distributed. Then if $rank(L) \leq \rho_rn\mu^{-1}(\log{n})^{-2}$ and $|\Omega| \leq \rho_s n^2$ with $\rho_r,\rho_s > 0$ being sufficiently small numerical constants. With high probability, the solution $(\hat{L},\hat{S})$ to (\ref{LSD}) satisfies:
%  \begin{equation}
%  ||\hat{L} - L ||_F^2 + ||\hat{S} - S||_F^2 \leq Cn^2\delta^2,
%  \end{equation}
%   where C is a numerical constant, and $n= \max(N,T)$.
%  \end{theorem}
%
%  \begin{lemma}
%  Assume $||\mathcal{P}_{\Omega} \mathcal{P}_{\mathcal{T}} ||\leq 1/2, \lambda \leq 1/2$, let $\hat{X} = \langle \hat{L}, \hat{S} \rangle$ be the solution to problem (\ref{LSD}), then $\hat{X}$ satisfies:
%  \begin{equation}
%  ||\hat{X} - X ||_F = ||\hat{L}+\hat{S}-L-S||_F \leq (8n\sqrt{40n + 40n/p + 5}+ \sqrt{2})\delta.
%  \end{equation}
%  \end{lemma}

  \begin{theorem}\label{Main_results}
  Suppose that $L$ obeys (\ref{LS_condition}) and the support of $S$ is uniformly distributed. Then if $rank(L) \leq \rho_rn\mu^{-1}(\log{n})^{-2}$ and $|\Omega| \leq \rho_s n^2$ with $\rho_r,\rho_s > 0$ being sufficiently small numerical constants. Assume $||\mathcal{P}_{\Omega} \mathcal{P}_{\mathcal{T}} ||\leq 1/2, \lambda \leq 1/2$, let $\hat{X} = \langle \hat{L}, \hat{S} \rangle$ be the solution to (\ref{LSD}). Then with high probability, $\hat{X}$ satisfies:
  \begin{equation}
   ||\hat{X} - X ||_F = ||\hat{L}+\hat{S}-L-S||_F \leq (8n\sqrt{40n + 40n/p + 5}+ \sqrt{2})\delta,
  \end{equation}
  where $n= \max(N,T)$.
  \end{theorem}

\section{Revealing Anomaly Readings in Real-World Wireless Sensor Networks} \label{observation}

\subsection{Data Sets}

   Table I lists the sensory data matrices used in this paper. These data sets are collected by the IntelLab \cite{Intel}, GreenOrbs \cite{liu2011does}, and NBDC-CTD \cite{NBDC} projects, which are deployed in indoor, mountain and ocean environments, respectively.

   \textbf{IntelLab} \cite{Intel}: is deploy in the Intel Berkeley Research lab from Feb. 28th to Apr. 5th, 2004. There are 54 Mica2 nodes places in a 40m $\times$ 30m room. The nodes are set to send back a packet every 30 seconds. This data set includes in total $2,313,682$ data packets.

   \textbf{GreenOrbs} \cite{liu2011does}: is deployed on the Tianmu Mountain, Zhejiang Province, China. In total, there are 326 nodes deployed. Nodes are set to transmitting packets back to the sink node every minute. We use the data collected in Aug. 3nd $\sim $5th, 2011, including $308,928$ data packets.

   \textbf{NBDC CTD} \cite{NBDC}: by the National Oceanic and Atmospheric Administration's (NOAA) National Data Bouy Center (NDBC). CTD (Conductivity, Temperature, and Depth) is a shipboard device consisting of many small probes. Nodes are set to report every $10$ minutes. We use $8, 107$ data packets collected during Oct. 26th $\sim$ 28th, 2012.

\subsection{Data Loss Rate}

   Due to unreliable wireless communication and packet collision, the raw data matrices at the sink node are incomplete. To quantify the data loss problem, we define a loss rate for each node as
   $r_i = \mathcal{L}_i/T,~i\in \{1,2,...,n\}$, where $\mathcal{L}_i$ is the number of lost readings of the $i$-th sensor node. The empirical cumulative distribution function of data loss rates is shown in Fig. 4. The data loss rates in real-world sensor networks are quite high, being $5\% \sim 95\%$.

      \begin{table*} \label{regularity_table}\centering
   \setlength{\abovedisplayskip}{4pt}
   \caption{Data sets for the compressibility characterization.}
   \begin{tabular}{|c|c|c|c|c|c|c|}
   \hline
   Data Sets & Environment & Nodes & Time period & Resolution  & Physical conditions & Size\\
   \hline\hline
   IntelLab & Indoor & 54 & Feb.28 $\sim$ Apr.5, 2004 & 30 seconds & Temperature, light, humidity & ~54 $\times$ 500  \\
   \hline
   GreenOrbs & Forest & 326 & Aug.03 $\sim$ 05, 2011 & 60 seconds & Temperature, light, humidity& 326  $\times$  750  \\
   \hline
   NBDC CTD & Ocean & 216 & Oct.26 $\sim$ 28, 2012 & 10 minutes & Temperature, salt, conductivity & 216 $\times$ 300 \\
   \hline
   \end{tabular}\vspace{-6pt}
   \end{table*}
   
      \begin{figure*}
   \centering
   \subfigure[Intel Indoor]{\includegraphics[width=0.25\textwidth]{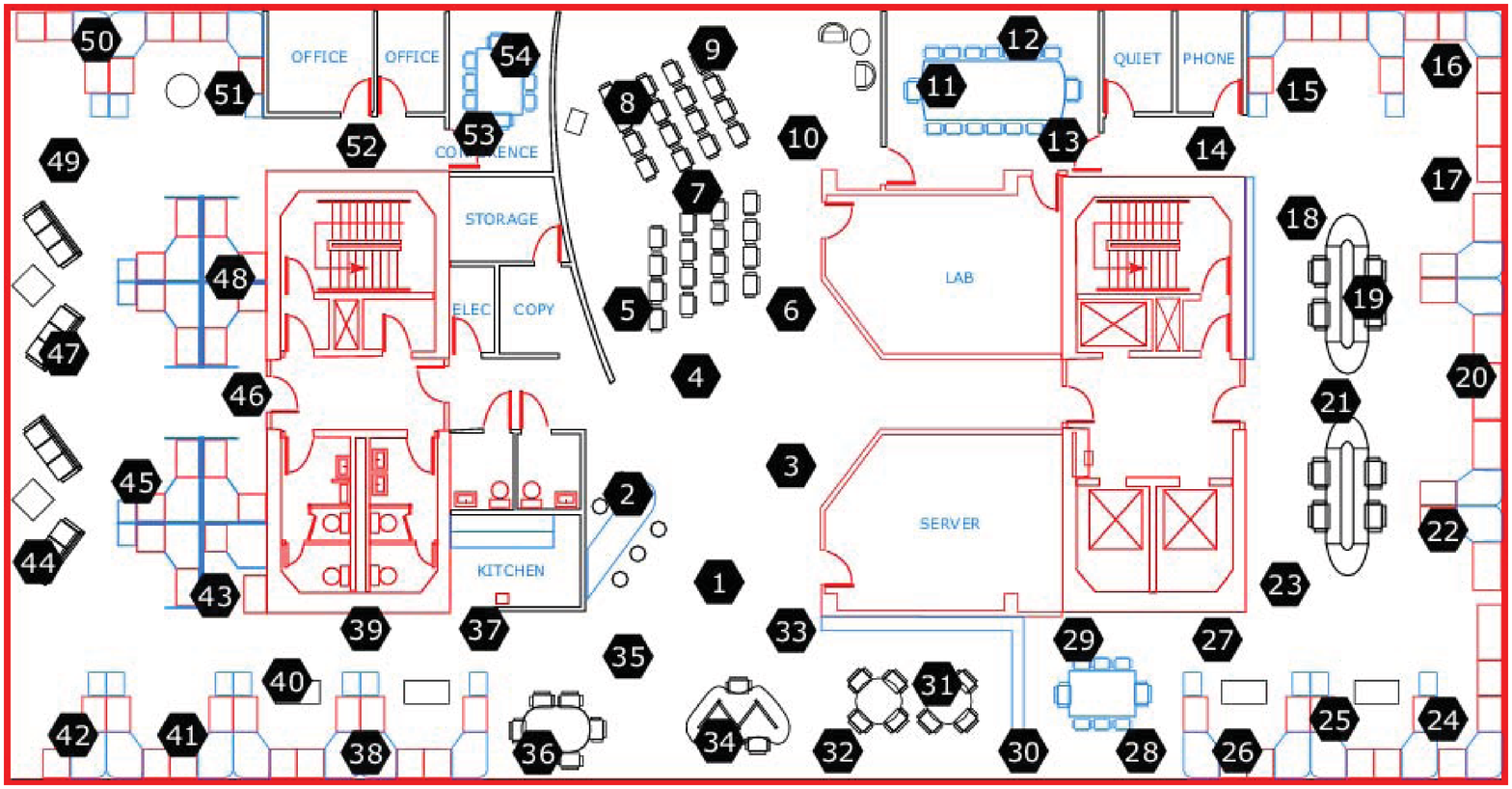}}
   \subfigure[GreenOrbs]{\includegraphics[width=0.40\textwidth]{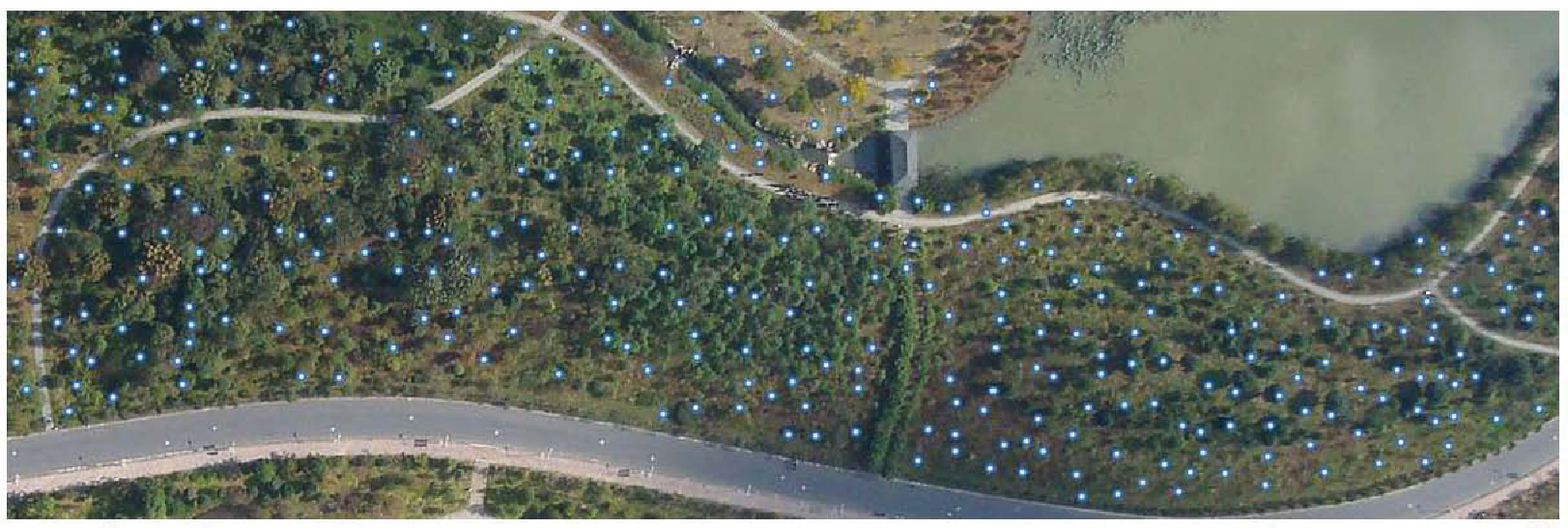}}
   \subfigure[NBDC CTD]{\includegraphics[width=0.33\textwidth]{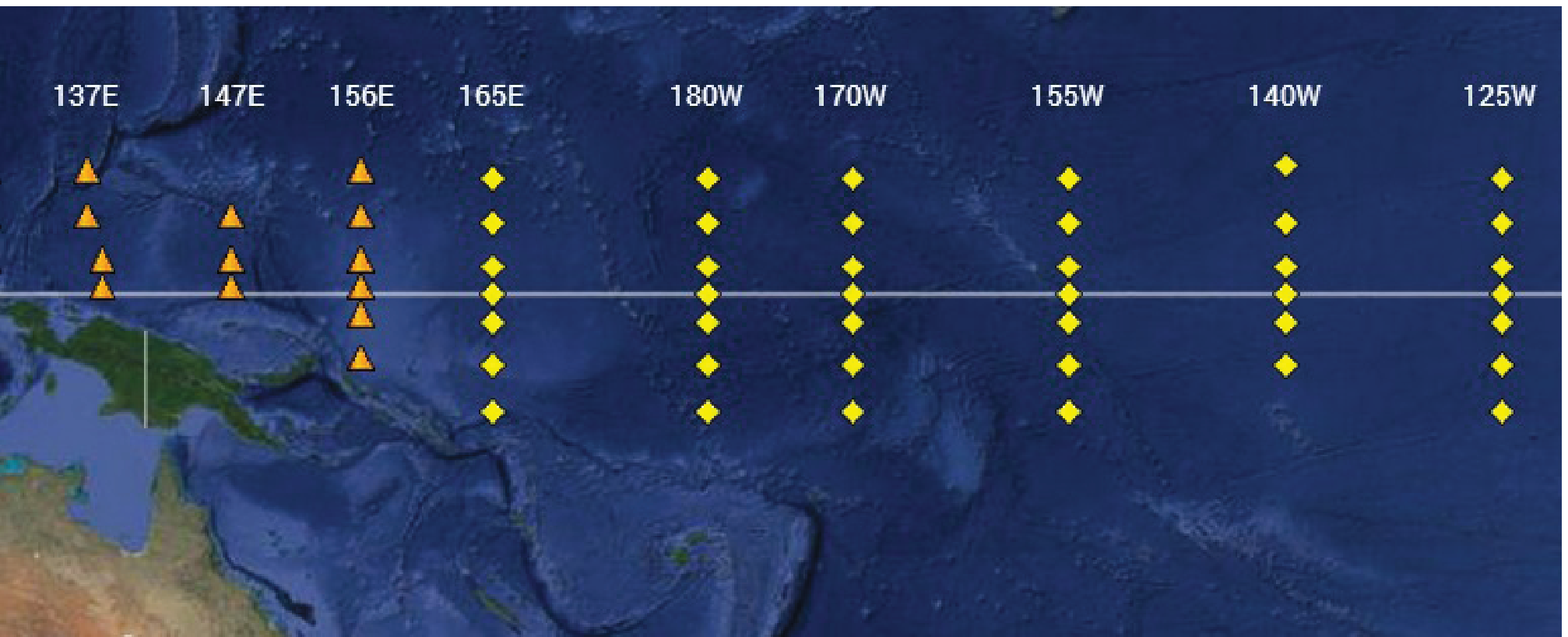}}
   \vspace{-8pt}
   \caption{Overview of the sensor network deployment for the Intel Indoor, GreenOrbs and NBDC CTD projects.}
   \vspace{-14pt}
   \end{figure*}

  \begin{figure*}
  \centering
  \subfigure{\includegraphics[width=0.42\textwidth]{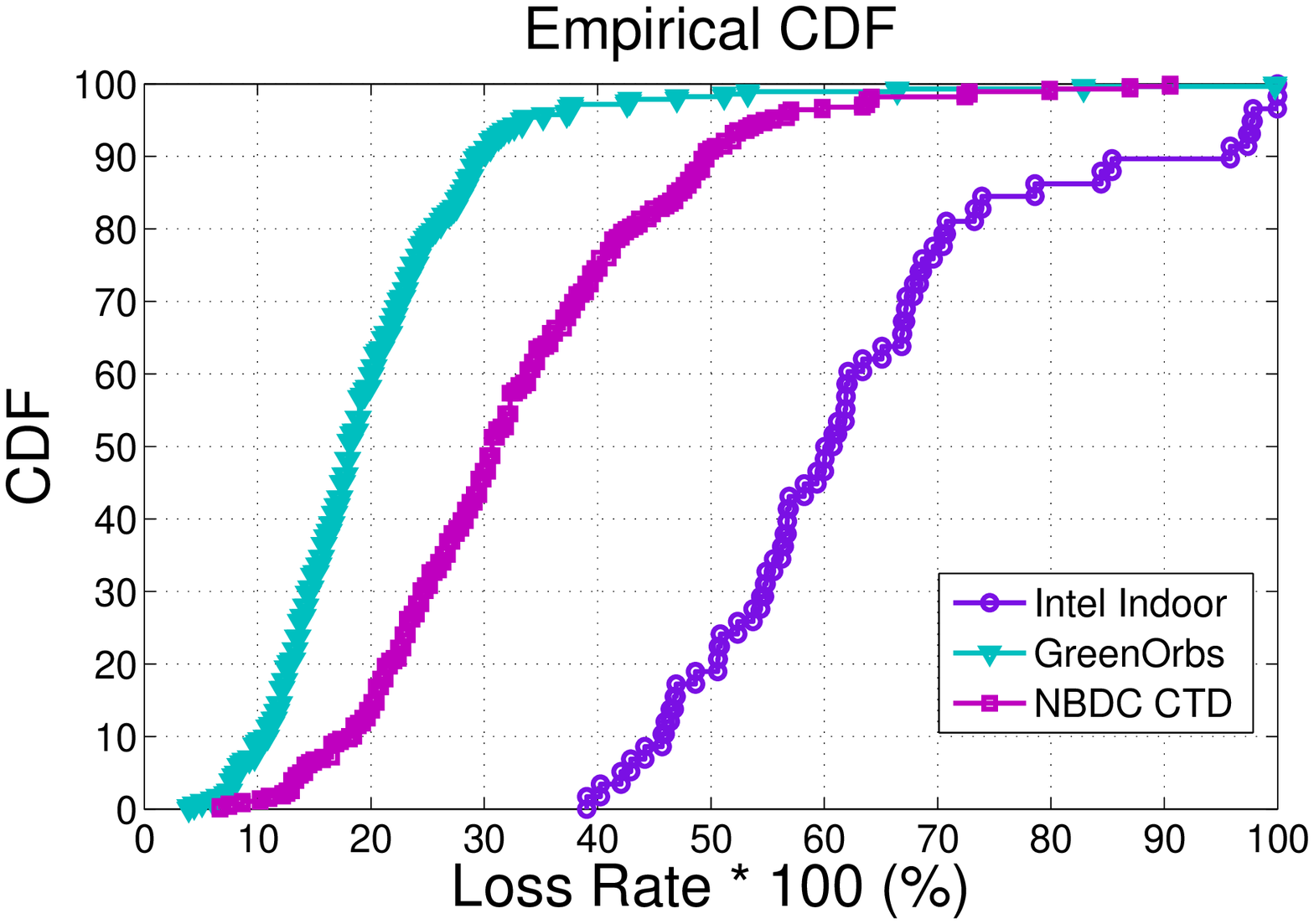}}
  \vspace{-3pt}
  \hspace{60pt}
  \subfigure{\includegraphics[width=0.42\textwidth]{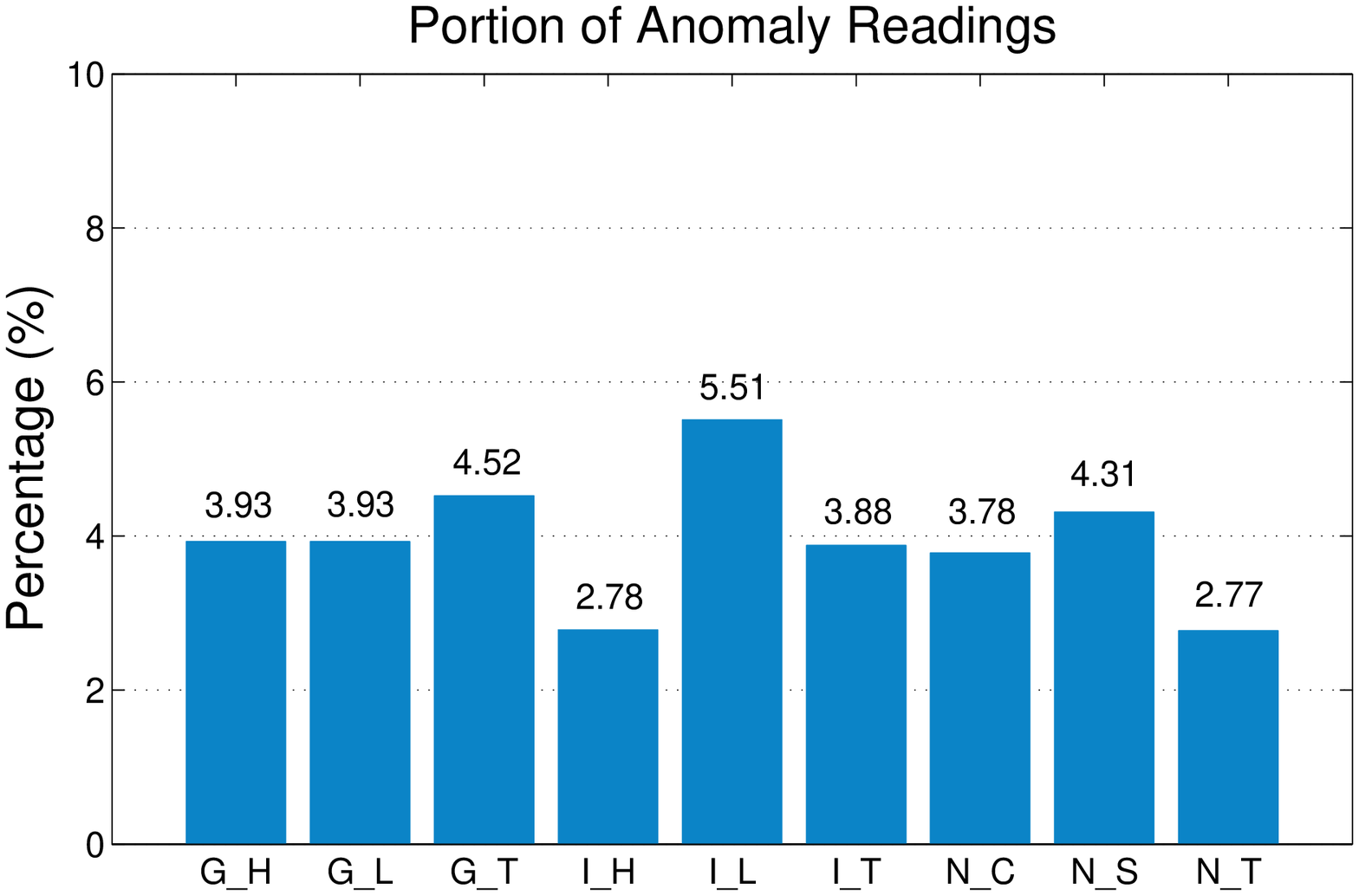}}
  \vspace{-8pt}
  \caption{(Left) The data loss rates of the IntelLab, GreenOrbs, and NBDC-CTD projects, deployed in indoor, mountain, and ocean environment, repectively. (Right) The percentage of anomaly readings, determined by the LS-Separation \cite{Wright2010Stable}. G,I,N are short for GreenOrbs, IntelLab, and NBDC CTD, while H, L, T, C, S are short for humidity, light, temperature, conductivity, and salt.}\label{fig:loss_rate}
  \vspace{-10pt}
  \end{figure*}

   \begin{figure*}
   \centering
   \subfigure{\includegraphics[width=0.33\textwidth]{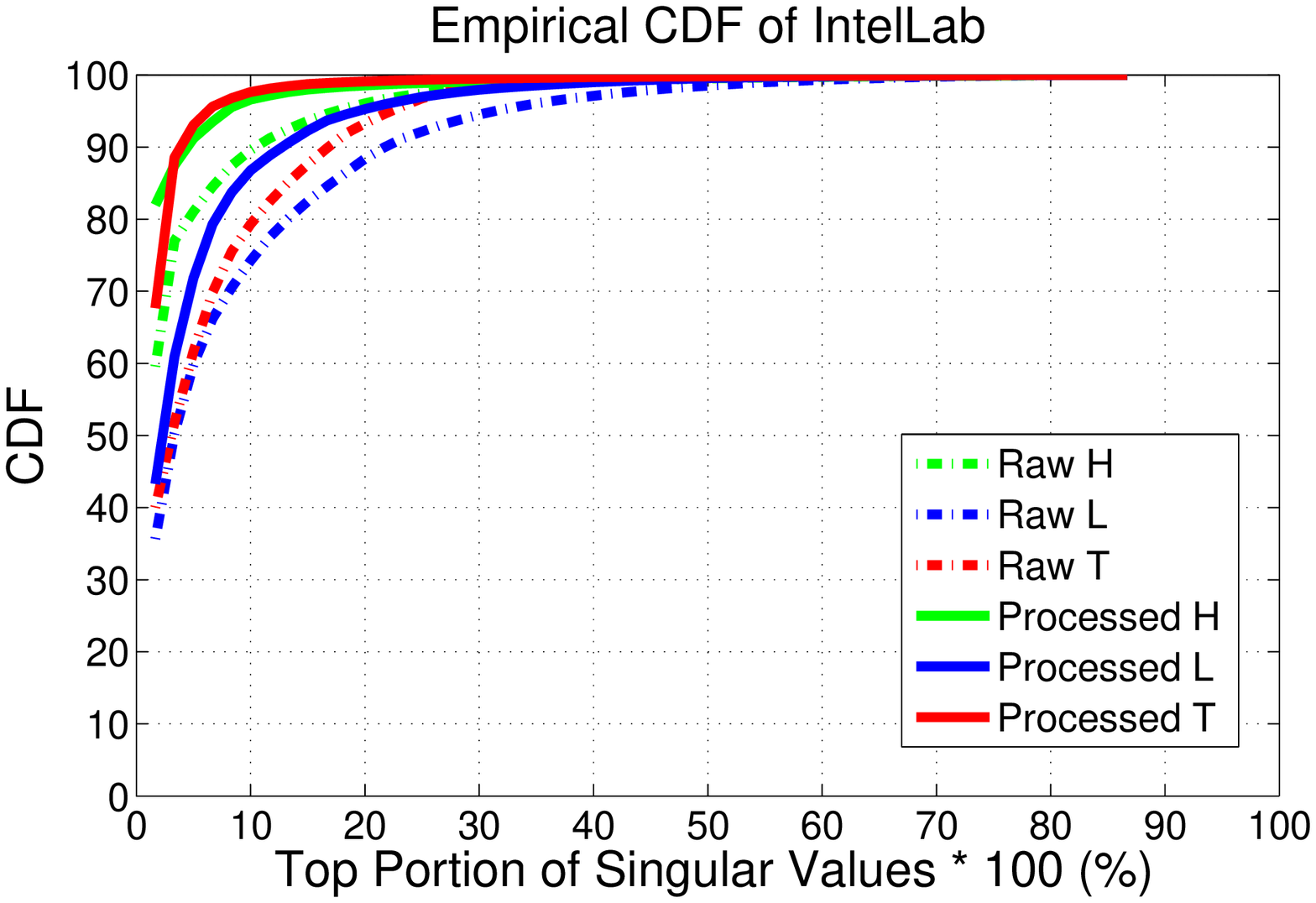}}
   \subfigure{\includegraphics[width=0.30\textwidth]{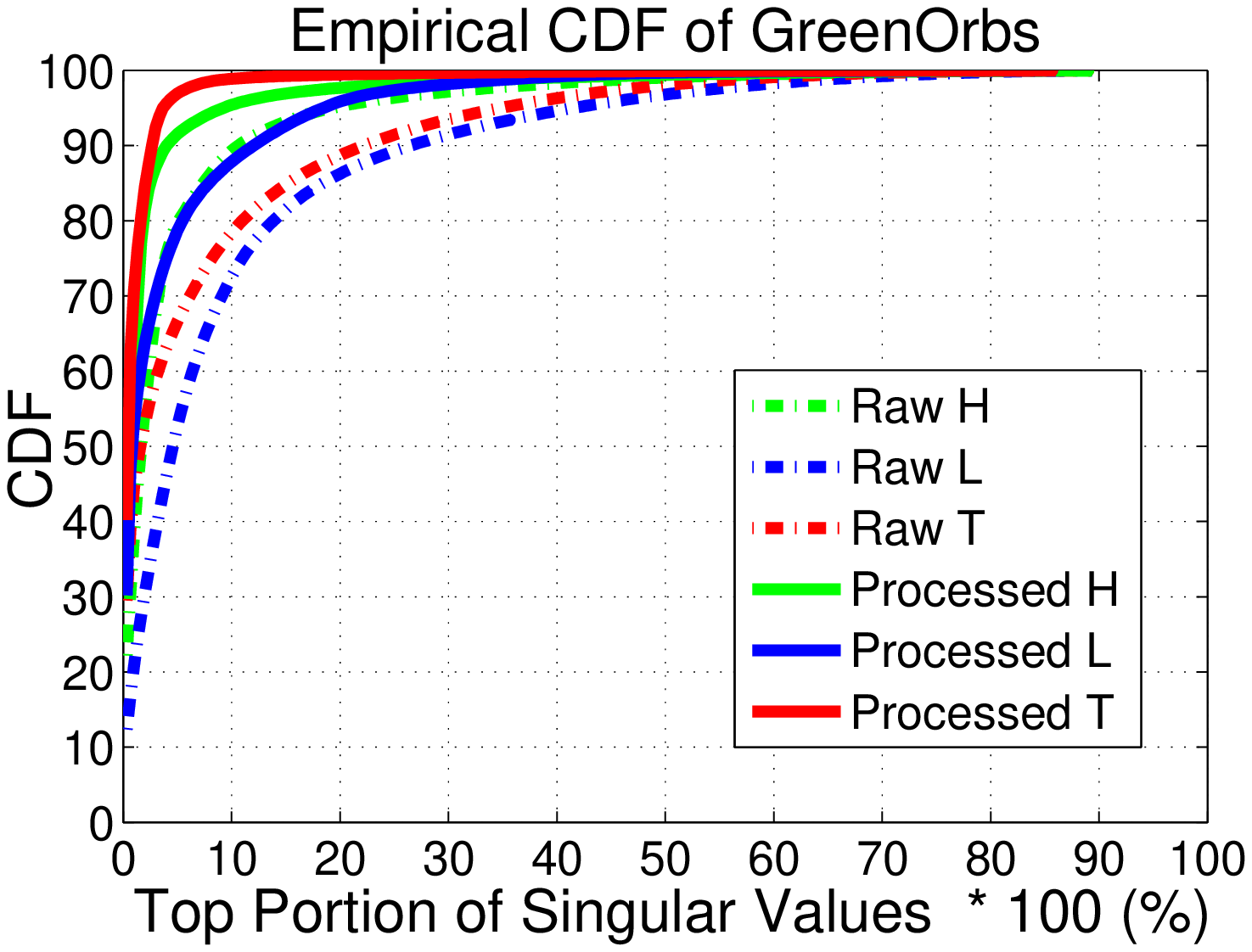}}
   \subfigure{\includegraphics[width=0.33\textwidth]{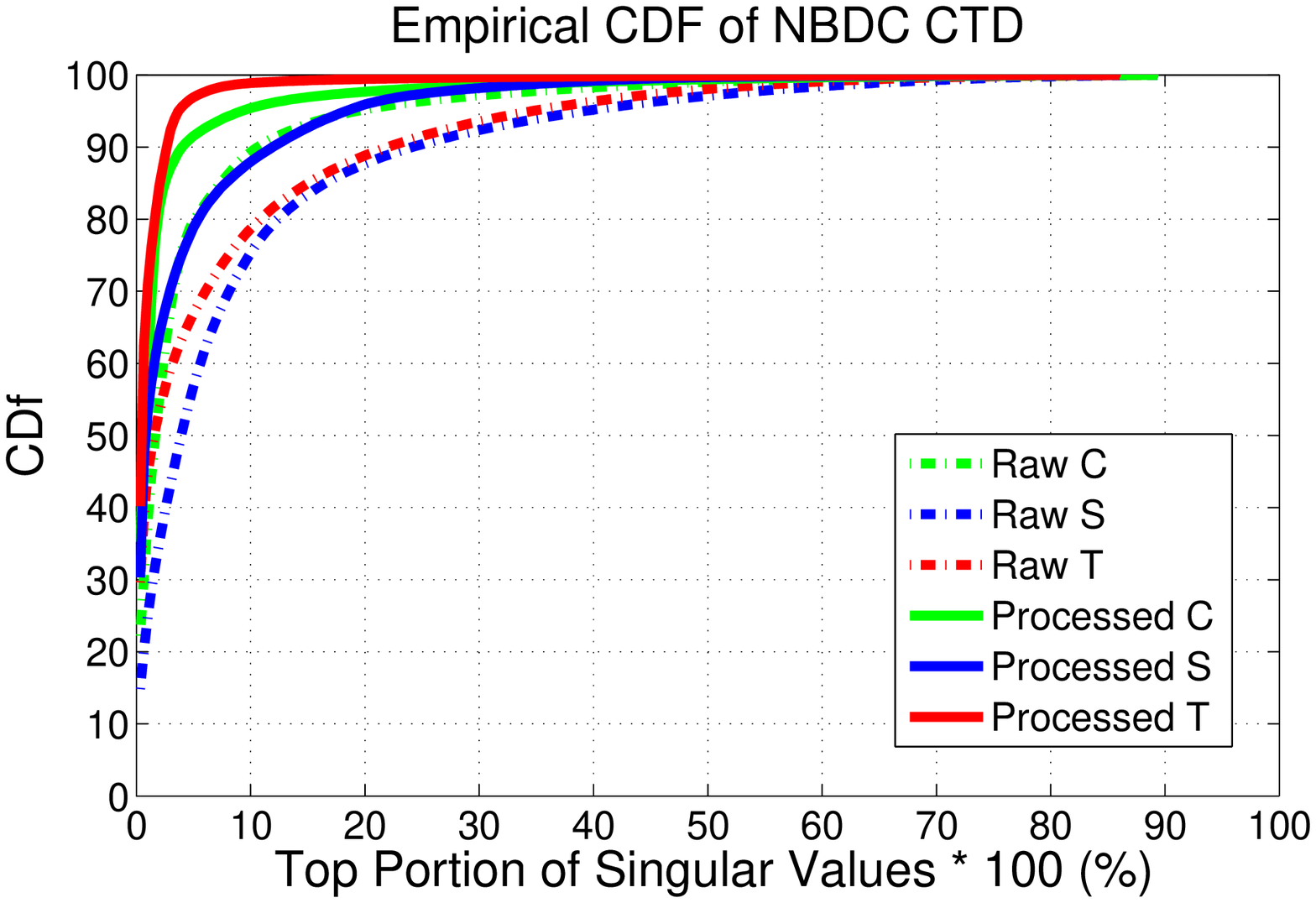}}
   \vspace{-8pt}
   \caption{The CDF of Singular Values for IntelLab, GreenOrbs, and NBDC CTD: svd of the raw data matrix, compared with processed data matrix. H, L, T, C, S are short for humidity, light, temperature, conductivity, and salt.}
   \label{fig:svd}
   \vspace{-14pt}
   \end{figure*}

\subsection{Low-Rank Property with Anomaly Readings}

   First, we construct full data matrices listed in Table. I, serving as the ground truth for observations and evaluations. Set a time window which is ten times the monitoring resolution for each data set. For each time slot, we randomly select one of the observed readings.

   For each raw data matrix, we apply singular value decomposition (SVD) to examine if it has a good low-rank approximation. The metric we use is the faction of total energy captured by the top portion of singular values. For example, the top $10\%$ portion of singular values for the IntelLab matrices means the largest $5$ singular values. We calculate $\left( \sum_{i=1}^{5} \sigma_i \right) / \left( \sum_{i=1}^{54} \sigma_i \right)$, where $\sigma_i$ is the $i$-th largest singular value and $\left( \sum_{i=1}^{54} \sigma_i \right)$ measures the total energy of the data matrix. Note that $1 - \left( \sum_{i=1}^{5} \sigma_i \right) / \left( \sum_{i=1}^{54} \sigma_i \right)$ is the relative approximation error of the best rank-5 approximation with respect to the Frobenius norm.

   Next we process the raw data matrices with {\em LS-separation} \cite{Wright2011}\cite{TFOCS2011}, which identifies the anomaly readings and returns a low-rank matrix. Then, we apply SVD on the processed low-rank matrices in a similar way.

   Fig. \ref{fig:loss_rate} shows the ratios of anomaly reading for each data matrix, ranging from $2.77\%$ to $5.51\%$. Fig. \ref{fig:svd} shows the CDFs of singular values for the raw data matrices and the processed data matrices. We can see that by excluding the anomalies, the processed data matrices has strengthened low-rank structures. For the IntelLab project, $18\%$ singular values capture $90\%$ energy of the raw temperature matrix while the ratio is reduced to $5\%$; for light, $22\%$ is reduced to $13\%$; and for humidity, $10\%$ is reduced to $5\%$. Similar observations hold for the GreenOrbs and NBDC CTD projects.

\section{Accelerated Proximal Gradient Approach}

\subsection{General Accelerated Proximal Gradient Algorithm}

  {\em Accelerated Proximal Gradient algorithms (APGs)} \cite{Proximal}, with the general flow shown in Table II, solve problems with the following general form:
  \begin{equation}\label{APG_problem_form}
  \setlength{\abovedisplayskip}{4pt}
  \min\limits_{X \in \mathcal{H}}~F(X)\triangleq \mu g(X) + f(X),
  \vspace{-4pt}
  \end{equation}
  where $\mathcal{H}$ is a real Hilbert space with a norm $||\cdot||$, $g$ is a continuous convex function, $f$ is convex and smooth with Lipschitz continuous gradient: $||\nabla f(X_1) - \nabla f(X_2)|| \leq L_f||X_1 - X_2||$ where $\nabla f$ is the Fre\'{e}chet derivative of $f$.

  APGs repeatedly minimize a proximal operator $Q(X,Y)$ to $F(X)$, defined as:
  \begin{equation*}\small
  \setlength{\abovedisplayskip}{4pt}
  Q(X,Y) \triangleq f(Y) + \langle \nabla f(Y), X-Y \rangle + \frac{L_f}{2}||X-Y||^2 + \mu g(X).
  \vspace{-4pt}
  \end{equation*}
  Since $f(X)$ is convex, for any $Y$, $Q(X,Y)$ upper bounds $F(X)$. If we define $G \triangleq Y - \frac{1}{L_f} \nabla f(Y)$, with simple transform and ignore the constant terms,
%  If we define $G \triangleq Y - \frac{1}{L_f} \nabla f(Y)$, since
%  \begin{equation*}\small
%  \begin{split}
%  &Q(X,Y) = f(Y) + \mu g(X) - \frac{L_f}{2} \langle \frac{1}{L_f} \nabla f(Y), \frac{1}{L_f} \nabla f(Y) \rangle + \frac{L_f}{2} \left[ \right.\\
%  & \langle \frac{1}{L_f} \nabla f(Y), \frac{1}{L_f} \nabla f(Y) \rangle + 2 \langle \frac{1}{L_f} \nabla f(Y), X-Y \rangle \\
%  &\left. + \langle X-Y, X-Y\rangle \right]\\
%  &= f(Y) + \mu g(X) - \frac{L_f}{2} \langle \frac{1}{L_f} \nabla f(Y), \frac{1}{L_f} \nabla f(Y) \rangle + \frac{L_f}{2} ||X-G||^2,
%  \end{split}
%  \end{equation*}
  then we have:
  \begin{equation}\small \label{Q_XY}
  \setlength{\abovedisplayskip}{4pt}
  \arg\min\limits_{X}~Q(X,Y) = \arg\min\limits_{X}~ \mu g(X) + \frac{L_f}{2} ||X-G||^2.
  \vspace{-4pt}
  \end{equation}
  We repeatedly set $X_{k+1} = \arg \min_X Q(X,Y_k)$, as in Table II. It is proved \cite{Converge} that setting $Y_k = X_k + \frac{t_{k-1}-1}{t_k}(X_k - X_{k-1})$ if $t_{k+1}^2 - t_{k+1} \leq t_k^2$ results in a convergence rate $O(1/k^2)$.

\begin{table}[t!]
\setlength{\abovedisplayskip}{6pt}
\caption{General Proximal Gradient Algorithm}
\vspace{-4pt}
\renewcommand{\arraystretch}{1.235}
\begin{tabular}{m{0.9\textwidth}}
\noalign{\hrule height 1.2pt}
1: \textbf{while} not converged \textbf{do} \\
       2:~~~~$Y_k \leftarrow X_k + \frac{t_{k-1}-1}{t_k}(X_k - X_{k-1})$;\\
       3:~~~~$G \leftarrow Y_k - \frac{1}{L_f} \nabla f(Y_k)$;\\
       4:~~~~$X_{k+1}\leftarrow \arg\min_{X}~{ \mu g(X) + \frac{L_f}{2} ||X-G_k||^2 }$;\\
       5:~~~~$t_{k+1} \leftarrow \frac{1+\sqrt{4t_{k}^2+1}}{2},~k\leftarrow k+1$;\\
6: \textbf{end while}\\
\noalign{\hrule height 1.2pt}
\end{tabular}
%\caption{Decoding Algorithm}
\label{t2}
\vspace{-6pt}
\end{table}

\begin{table}[t!]
\setlength{\abovedisplayskip}{6pt}
\caption{Accelerated Proximal Gradient Algorithm}
\vspace{-4pt}
\renewcommand{\arraystretch}{1.235}
\begin{tabular}{m{0.9\textwidth}}
\noalign{\hrule height 1.2pt}
\textbf{Input}: $M, \Omega, \lambda$.\\
1: $L_0,L_{-1} \leftarrow 0; S_0,S_{-1} \leftarrow 1; t_0,t_{-1} \leftarrow 1; \overline{\mu} \leftarrow \delta \mu_0$; k=0.\\
2: \textbf{while} not converged \textbf{do} \\
       3:~~~~$Y_k^L \leftarrow L_k + \frac{t_{k-1}-1}{t_k}(L_k - L_{k-1})$.\\
       4:~~~~$Y_k^S \leftarrow S_k + \frac{t_{k-1}-1}{t_k}(S_k - S_{k-1})$.\\
       5:~~~~$G_k^L \leftarrow Y_k^L - \frac{1}{2} \mathcal{P}_{\mathcal{O}}(Y_k^L + Y_k^S - R)$.\\
       6:~~~~$(U,S,V) \leftarrow svd(G_k^L), ~L_k = U \mathcal{S}_{\frac{\mu_k}{2}}[S]V^{*} $.\\
       7:~~~~$G_k^S \leftarrow Y_k^S - \frac{1}{2} \mathcal{P}_{\mathcal{O}}(Y_k^L + Y_k^S - R)$.\\
       8:~~~~$S_{k+1} \leftarrow \mathcal{S}_{\frac{\lambda \mu}{2}}[G_k^{S}]$.\\
       9:~~~~$t_{k+1} \leftarrow \frac{1+\sqrt{4t_{k}^2+1}}{2},~\mu_{k+1} \leftarrow max(\mu_k,\overline{\mu}),~k\leftarrow k+1$.\\
10: \textbf{end while}\\
\textbf{Output}: $\hat{L} \leftarrow L_k, ~\hat{S} \leftarrow S_k$.\\
\noalign{\hrule height 1.2pt}
\end{tabular}
%\caption{Decoding Algorithm}
\label{t2}
\vspace{-10pt}
\end{table}

\subsection{Robust Data Recovery by Accelerated Proximal Gradient}

  Consider the augmented Lagrangian function of (\ref{LSD}),
  \begin{equation}
  \arg \min\limits_{X} F(X) \triangleq \mu ||L||_{*} + \mu \lambda ||S||_1 + \frac{1}{2} ||\mathcal{P}_{\mathcal{O}}(R - L - S)||_F^2.
  \end{equation}

  Different from the conventional APGs, $\mathcal{H}$ is the space of same-sized matrices endowed with the Frobenius norm $||\cdot||_F$, our iterates $X_k$ are ordered pairs $(L_k,S_k) \in \mathbb{R}^{m \times n} \times \mathbb{R}^{m \times n}$, $g(X_k) = \mu ||L_k||_{*} + \mu \lambda ||S_k||_1$, and the Lipschitz constant $L_f = 2$.

  Write $G_k \triangleq (G_k^{L}, G_k^{S}) \in \mathbb{R}^{m \times n} \times \mathbb{R}^{m \times n}$, and let $USV^{*}$ be the singular value decomposition of $G_k^{L}$. Following the framework of APGs, we have:
  \begin{equation}
  \setlength{\abovedisplayskip}{4pt}
  L_{k+1} = U \mathcal{S}_{\frac{\mu}{2}}[S]V^{*}, ~~S_{k+1} = \mathcal{S}_{\frac{\lambda \mu}{2}}[G_k^{S}],
  \vspace{-4pt}
  \end{equation}
  where $\mathcal{S}[\cdot]$ is an element-wise threshold operator
  %\footnote{For vector or matrix, it operates on its element.}
  , defined as:
  \begin{eqnarray}\label{S}
  \setlength{\abovedisplayskip}{4pt}
  \mathcal{S}_{\epsilon}[x]=
  \begin{cases}
  x - \epsilon,&\text{if}~~x > \epsilon\cr
  x + \epsilon, &\text{if}~~x < -\epsilon\cr
  0,&\text{otherwise}\cr
  \end{cases}.
  \vspace{-4pt}
  \end{eqnarray}
  Therefore, we construct our accelerated proximal algorithm as in Table III. Please refer to the appendix for the proofs of the following two theorems.

\begin{theorem}\label{main_theorem}
 Let $F(X) \triangleq F(L,S) \triangleq \mu ||L||_{*} + \mu \lambda ||S||_1 + \frac{1}{2} ||\mathcal{P}_{\mathcal{O}}(R - L - S)||_F^2$. Then, for all $k \geq 1$, 
 %$k \geq k_0 \triangleq \frac{\log(\mu_0 / \mu)}{\log(1/\eta)}$, 
 our algorithm achieves the following converge rate:
 \begin{equation}
 \setlength{\abovedisplayskip}{4pt}
 F(X_k) - F(X^*) \leq \frac{4||X_{1} - X^* ||_F^2}{k^2},
 \vspace{-4pt}
 \end{equation}
 where $X^*$ is optimal solution to (\ref{LSD}).
\end{theorem}

\begin{lemma}\label{main_lemma}
 Let $F(X) \triangleq F(L,S) \triangleq \mu ||L||_{*} + \mu \lambda ||S||_1 + \frac{1}{2} ||\mathcal{P}_{\mathcal{O}}(R - L - S)||_F^2$. Our algorithm converges to its global optimal. To achieve $F(X_k)  - F(X^{*}) \leq \varepsilon$, our algorithm requires $k = O(1/\sqrt{\varepsilon})$ iterations.
\end{lemma}

\section{Evaluation}

\subsection{Evaluation Methodology}

  Our evaluations are based on the real-world data sets from the IntelLab \cite{Intel}, GreenOrbs \cite{liu2011does}, and NBDC-CTD \cite{NBDC} projects. The sensors in each project measure three types of physical conditions. Given the ground truth data described in Table I, we randomly drop entries and then compare the recovered data matrices with the ground truth. Under different data loss rate, we measure the recovery errors in terms of {\em Normalized Square Error (NSE)}, defined as following:
  \begin{equation}
  \setlength{\abovedisplayskip}{4pt}
  NSE = \frac{\sum_{i=1}^{N}\sum_{j=1}^{T} (\hat{L}_{ij} - L_{ij} )^2 }{\sum_{i=1}^{N}\sum_{j=1}^{T} L_{ij}^2 },
  \vspace{-4pt}
  \end{equation}
  where $\hat{L}$ is the estimated low-rank matrix, $L$ is the ground truth. $NSE$ is widely used in data interpolation and parameter estimation.

  We compare our LS-Decomposition with the standard matrix completion and the smooth-regulated matrix completion. To further understand the gap between LS-Decomposition and the optimal recovery performance, we construct an oracle solution which gives us information about the support $\Omega$ of $S$ and the row (column) space $\mathcal{T}$ of $L$.

  (1) Oracle Solution (OS) \cite{Wright2010Stable} serves as the optimal solution. Assume that we know the support $\Omega$ of $S$ and the row and column spaces $\mathcal{T}$ of $L$, which can be inferred by performing {\em LS-Separation} (problem P\ref{LS-Separation} as done in Section V) on the ground truth data matrices. It estimates $L$ and $S$ as the solution $L_{oracle}$ and $S_{oracle}$ to the following lease squares problem:
   \begin{equation}
   \begin{split}
   \setlength{\abovedisplayskip}{4pt}
   \langle L_{oracle}, S_{oracle} \rangle  &= \min\limits_{\langle L, S\rangle}~~||R - L - S||_F, \\
   &s.t.~~L \in \mathcal{T}, S \in \Omega.
   \vspace{-4pt}
   \end{split}
   \end{equation}

   (2) Matrix Completion (MC) \cite{Candy2009}\cite{Tao2010}\cite{TFOCS2011}: the standard matrix completion solves the following optimization problem:
   \begin{equation}
   \begin{split}
   \setlength{\abovedisplayskip}{4pt}
   \hat{L}  &= \min\limits_{L}~~||L||_{*} , \\
   &s.t.~~||R - L||_F \leq \delta.
   \vspace{-4pt}
   \end{split}
   \end{equation}

   (3) SRMF \cite{Kong2013INFOCOM}: it uses Sparsity Regularized Matrix Factorization that leverages both low-rank and spatio-temporal characteristics, as follows:
   \begin{equation}
   \begin{split}
   \setlength{\abovedisplayskip}{4pt}
   \hat{L}  &= \min\limits_{L}~~||L||_{*} + \lambda \mathcal{S}(L) , \\
   &s.t.~~||R - L||_F \leq \delta,
   \vspace{-4pt}
   \end{split}
   \end{equation}
   where $\mathcal{S}(\cdot)$ is the smooth term, and $\lambda$ is the weight to balance the low-rank term and the smooth term, which is usually determined by experiments. We find $\lambda = 0.01$ gives the best performance in our simulations. $\mathcal{S}(L)$ is defined via the diversity of matrix horizontal and vertical difference:
   \begin{equation}
   \setlength{\abovedisplayskip}{4pt}
   \mathcal{S}(L) = || \mathcal{D}_x (L)||_F^2 + || \mathcal{D}_y (L)||_F^2,
   \vspace{-4pt}
   \end{equation}
   where $\mathcal{D}_x (L)$ is an $N \times (T-1)$ matrix representing the horizontal difference of $L$ with element in form of $\mathcal{D}_x (i,j) = L(i, j+1) - L(i,j)$, and $\mathcal{D}_x (L)$ is an $(N-1) \times T$ matrix representing the horizontal difference of $L$ with element in form of $\mathcal{D}_y (i,j) = L(i+1, j) - L(i,j)$.

%
%   For completeness, we also quantify the anomaly detection performance on the observed entries, i.e., the fraction of real anomalies (only on the subset $\Omega \cap \mathcal{O}$) that are correctly identified. However, the MC, SRMF do not have the capability of detecting anomalies, thus for detection we only compare LS-Decomposition with the oracle solution.

\subsection{Performance Results}
   \begin{figure*}
   \centering
   \subfigure{\includegraphics[width=0.32\textwidth]{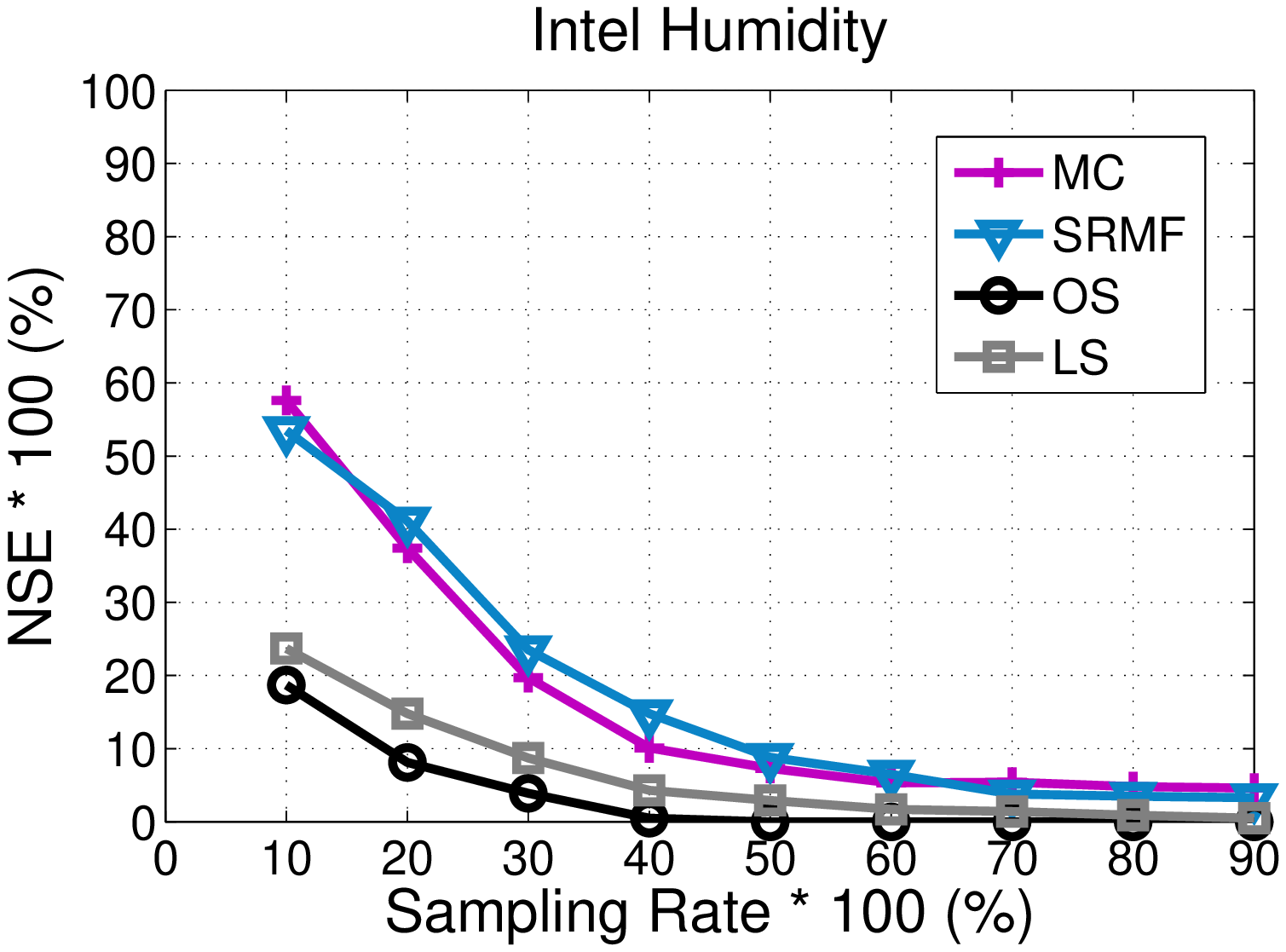}}
   \subfigure{\includegraphics[width=0.32\textwidth]{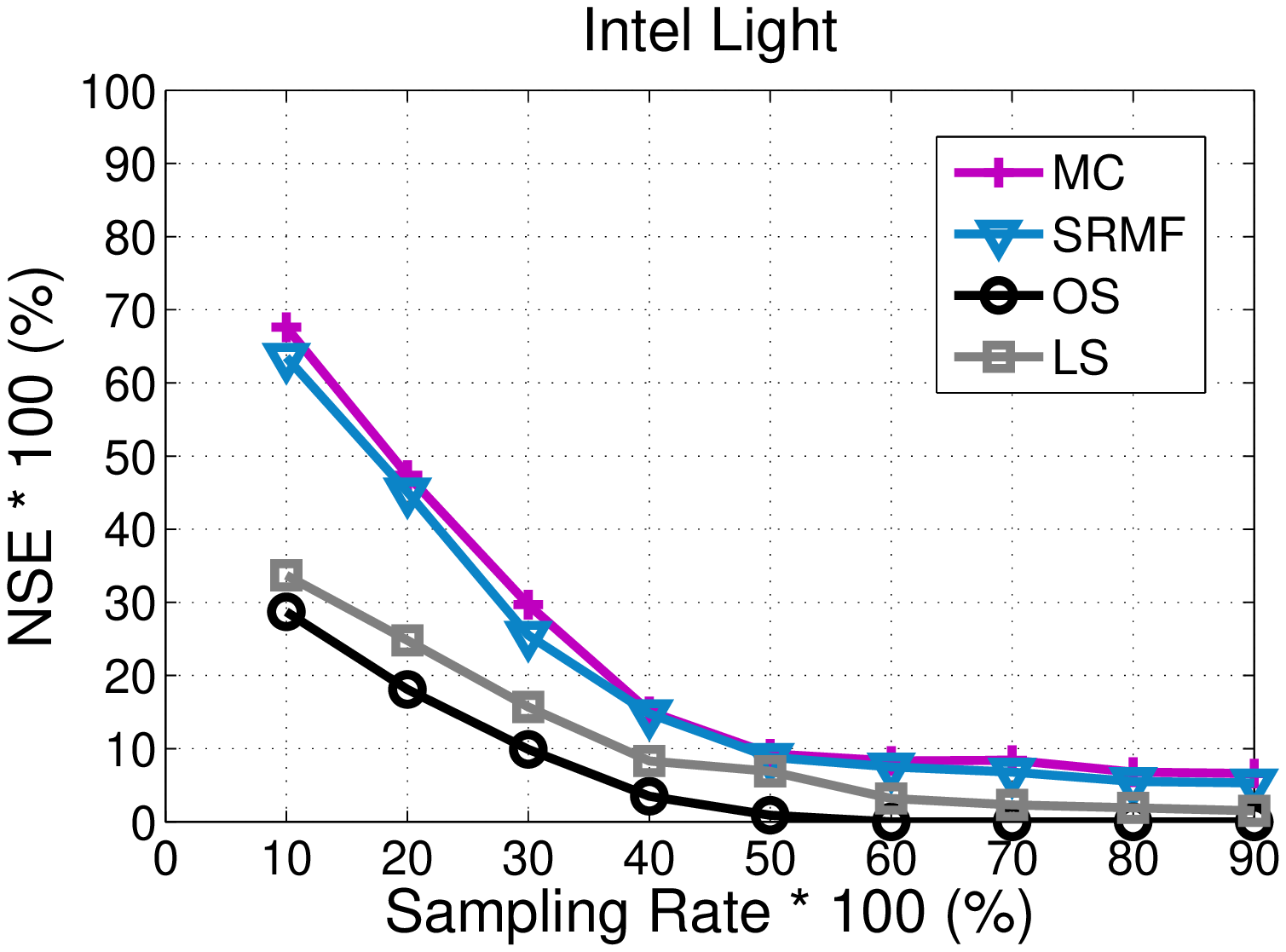}}
   \subfigure{\includegraphics[width=0.32\textwidth]{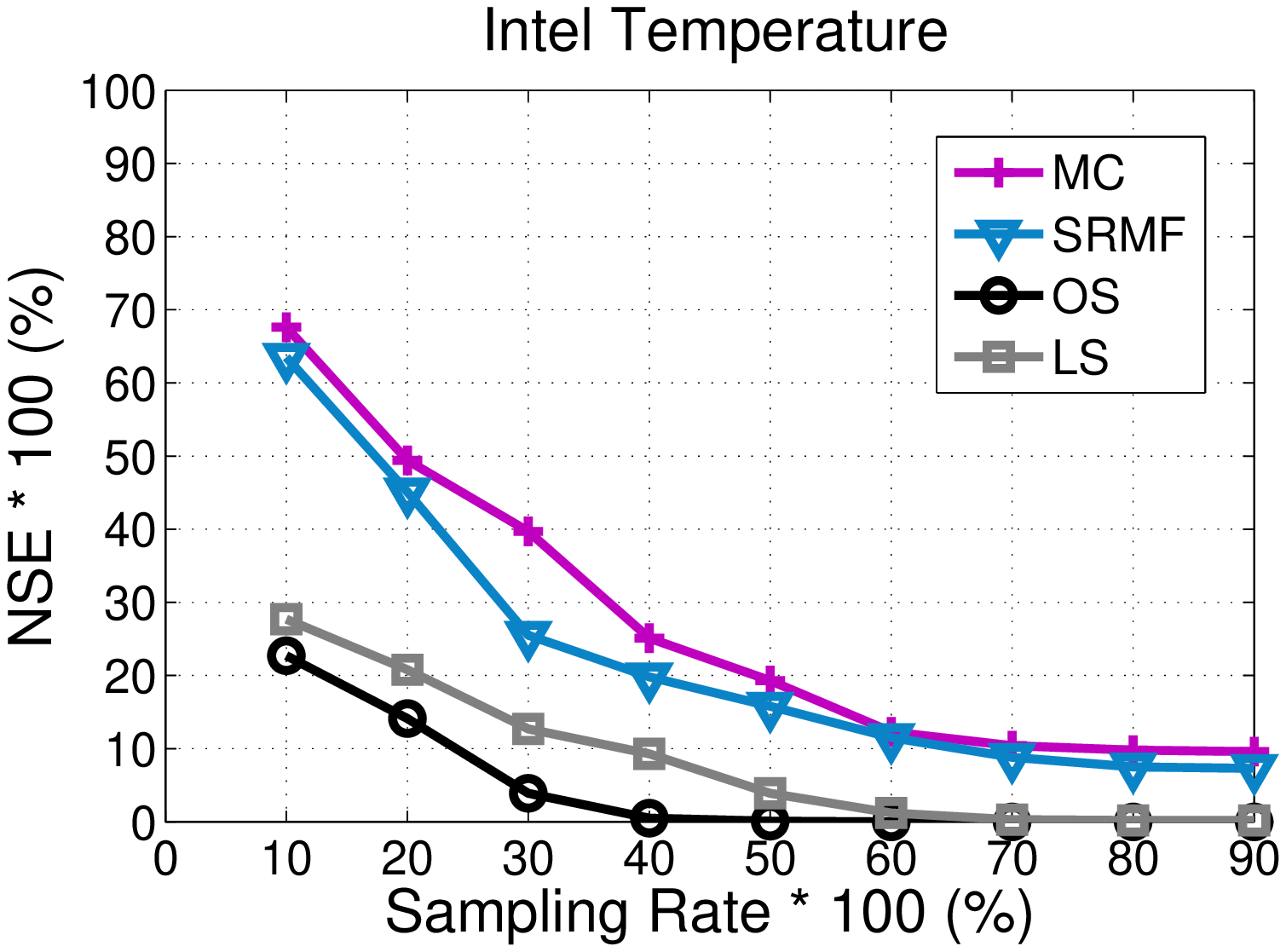}}
   \vspace{-6pt}
   \caption{Comparisons for humidity, light, temperature of the IntelLab project.}
   \vspace{-10pt}
   \end{figure*}

   \begin{figure*}
   \centering
   \subfigure{\includegraphics[width=0.32\textwidth]{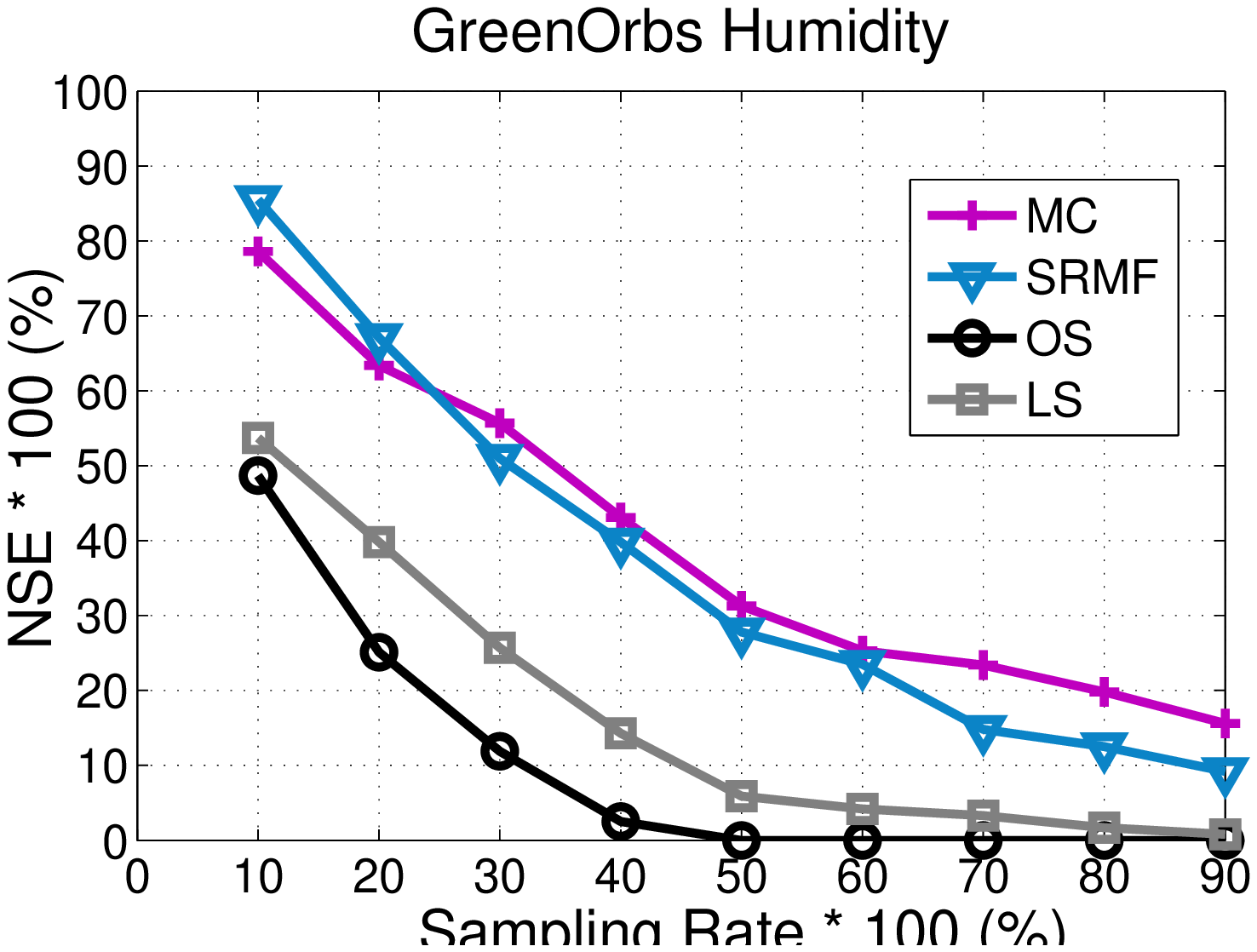}}
   \subfigure{\includegraphics[width=0.32\textwidth]{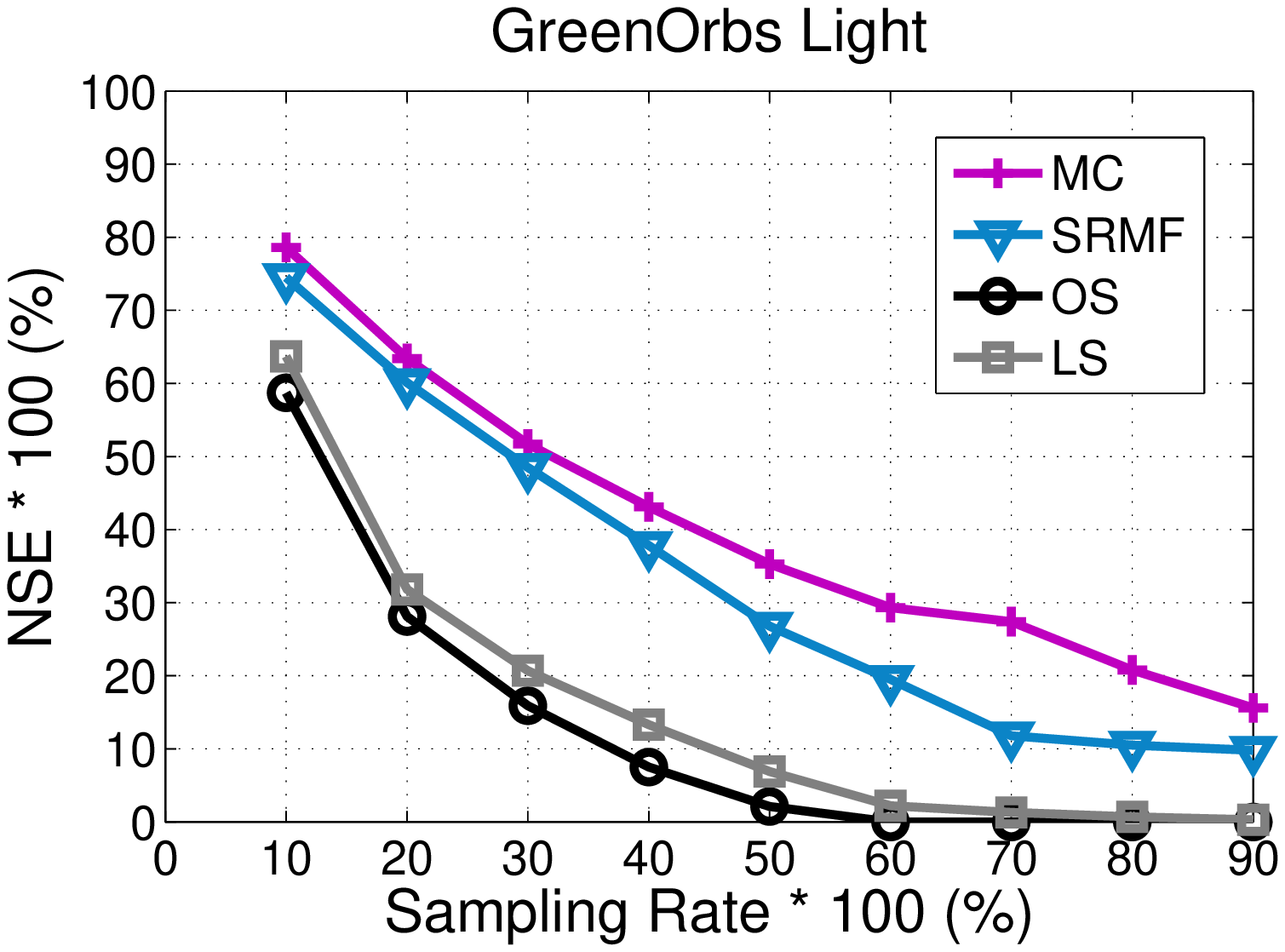}}
   \subfigure{\includegraphics[width=0.32\textwidth]{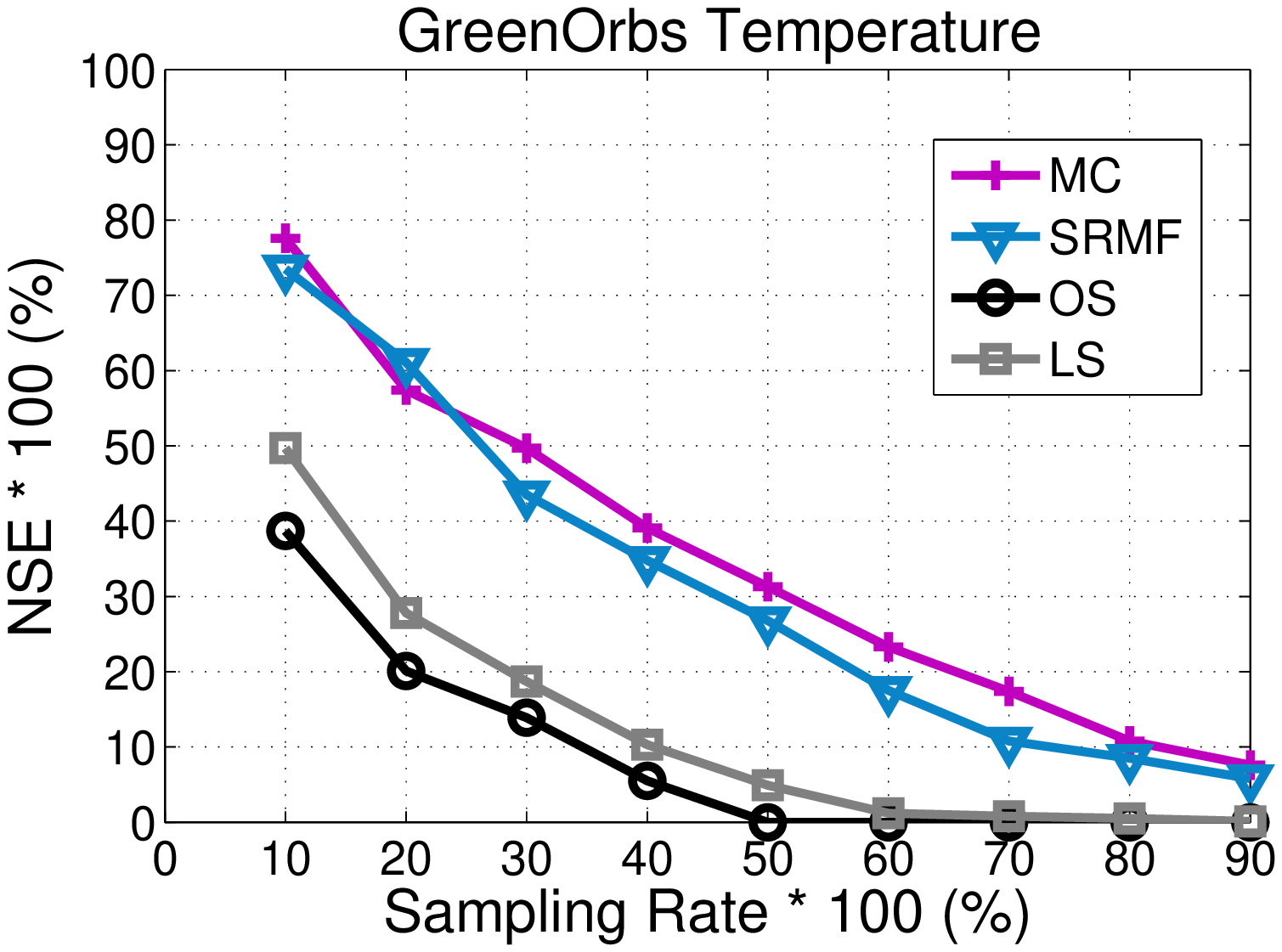}}
   \vspace{-6pt}
   \caption{Comparisons for humidity, light, temperature of the GreenOrbs project.}
   \vspace{-10pt}
   \end{figure*}

   \begin{figure*}
   \centering
   \subfigure{\includegraphics[width=0.32\textwidth]{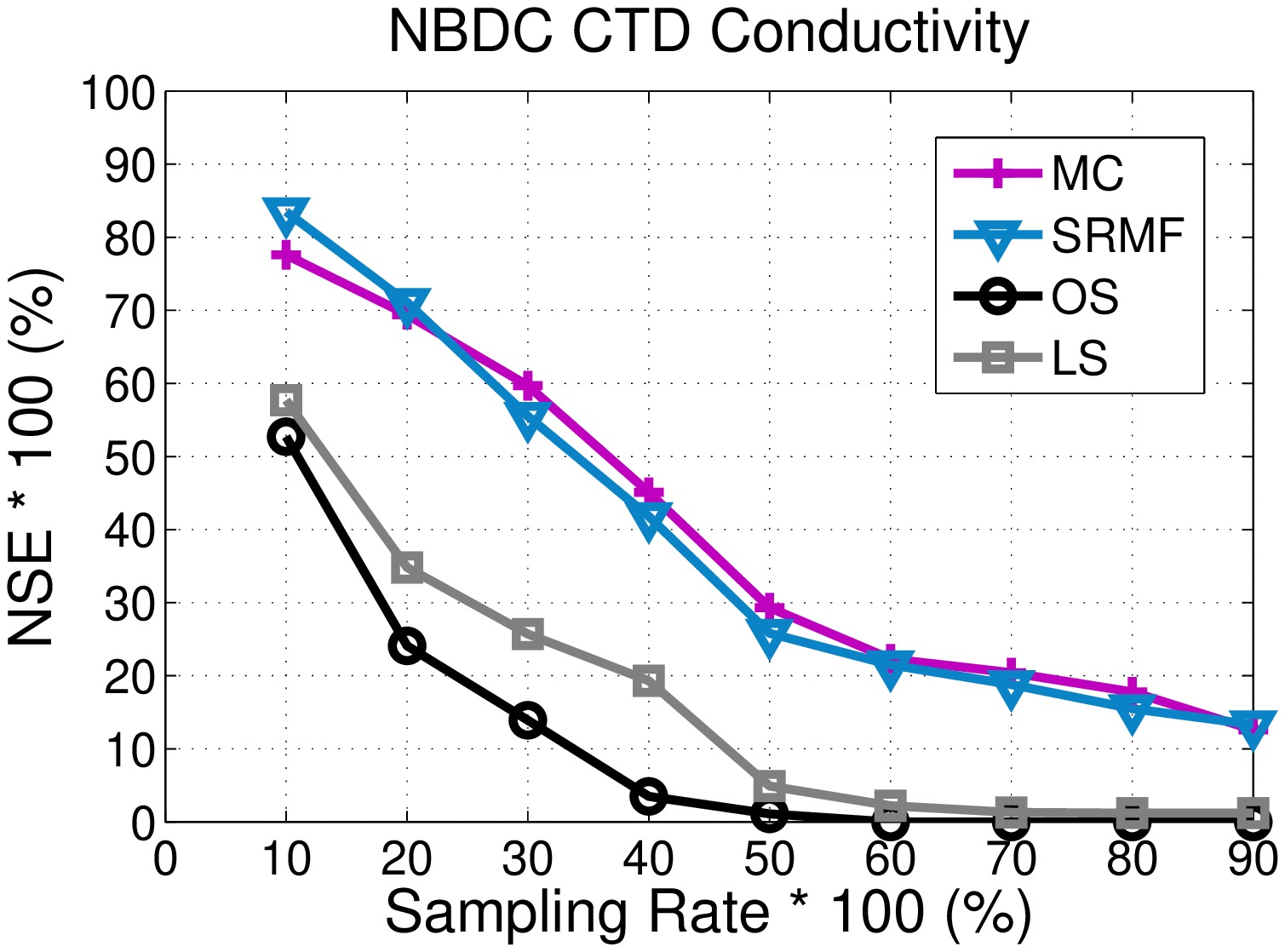}}
   \subfigure{\includegraphics[width=0.32\textwidth]{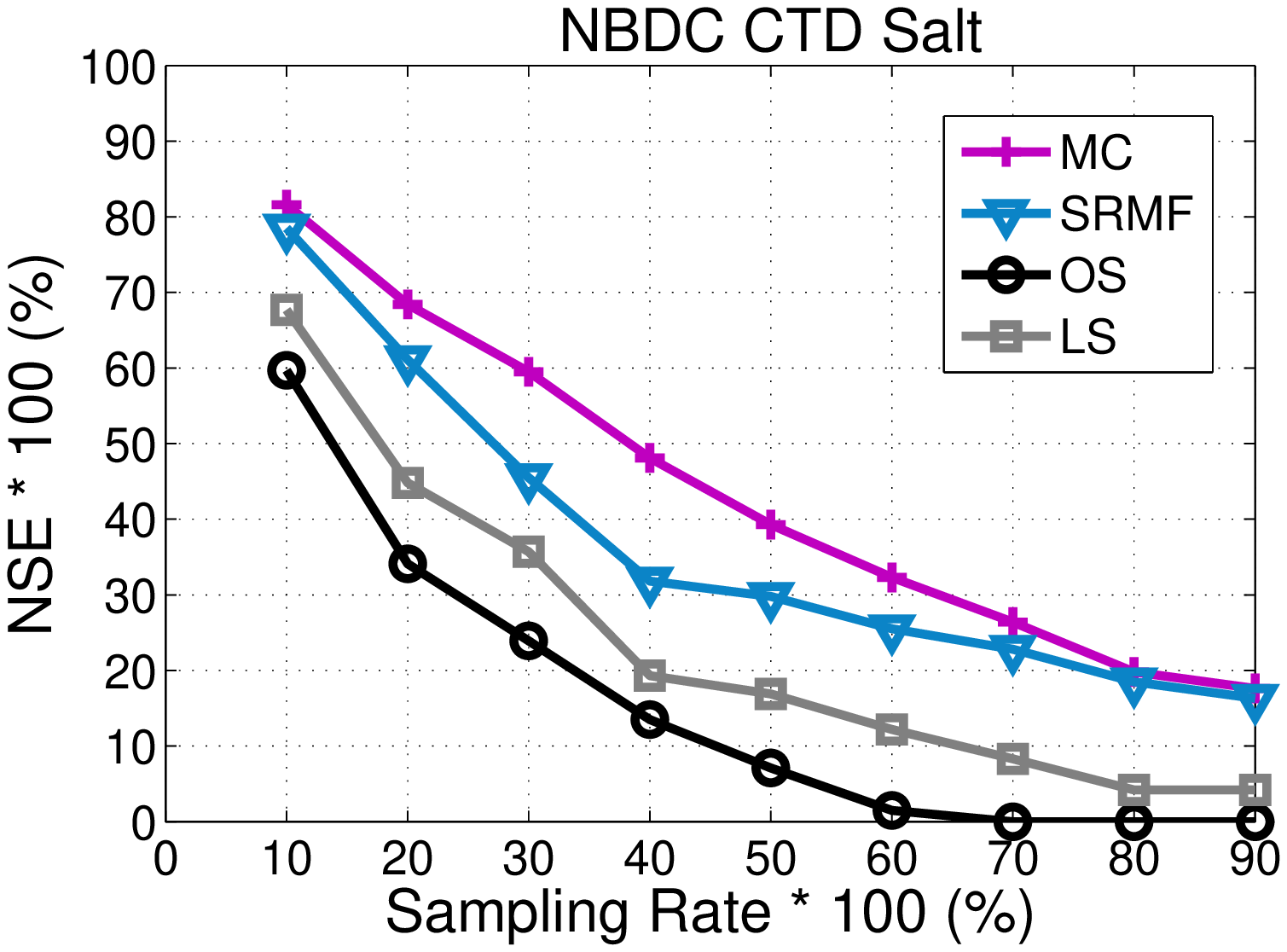}}
   \subfigure{\includegraphics[width=0.32\textwidth]{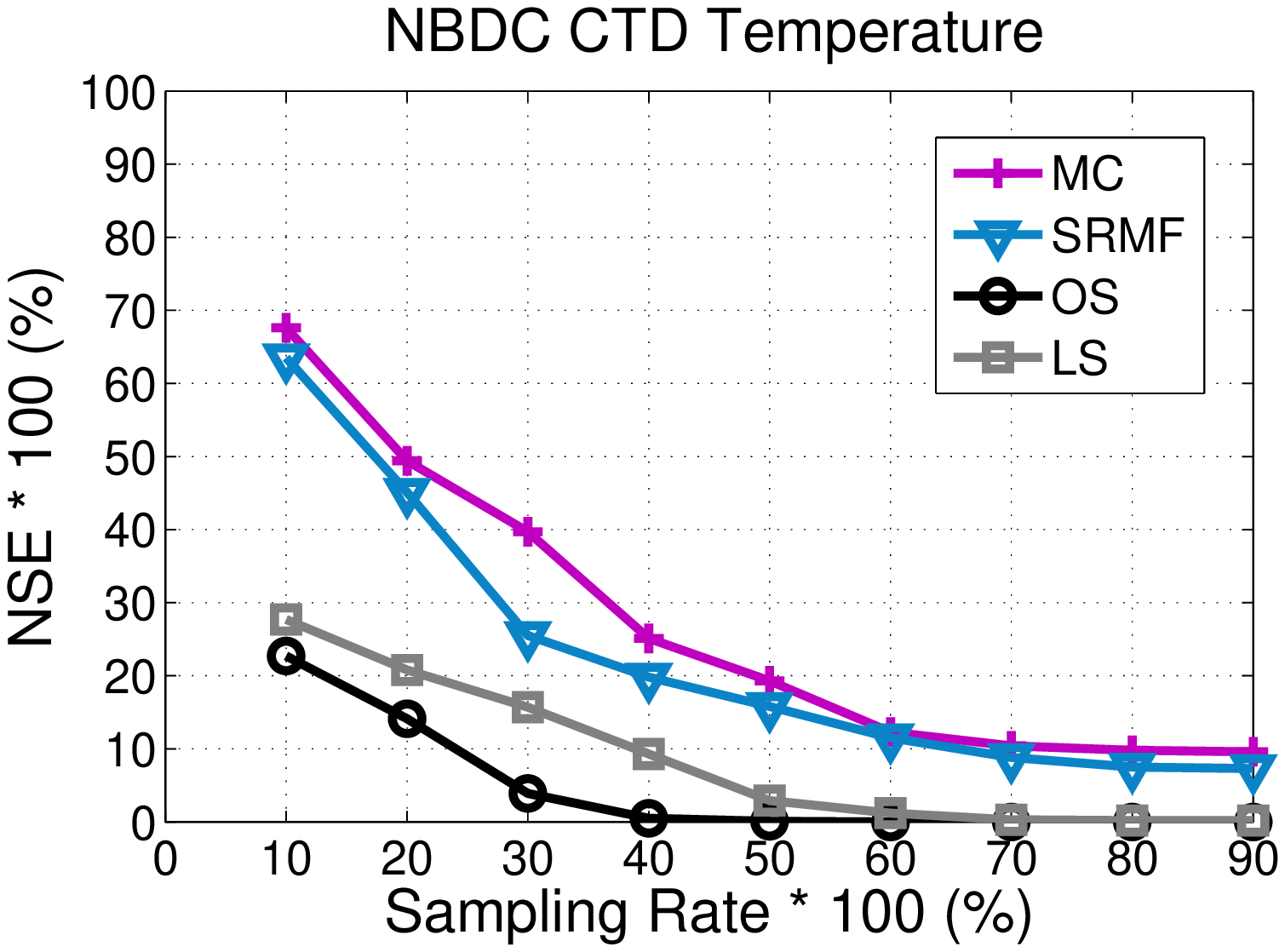}}
   \vspace{-6pt}
   \caption{Comparisons for conductivity, salt, temperature of the NBDC CTD project.}
   \vspace{-10pt}
   \end{figure*}

   We randomly drop entries with data loss rate $10\%$, $20\%$, ..., $90\%$, i.e., with sampling rate $90\%$, $80\%$, ..., $10\%$. Using the remained data as input, we apply the above four schemes to perform data recovery. For the IntelLab, GreenOrbs, and NBDC CTD projects, we test each physical attributes separately. We conduct 10 runs for each case and report an average over these 10 runs. %, as shown in Fig. 6, 7, and 8.

   From Fig. 6, 7, and 8, as the sampling rate increases (or the data loss rate decreases), the recovery errors for all the four algorithms decrease. LS performs much better than MC and SRMF which do not take anomaly readings into consideration. Generally, our scheme achieves recovery error $\leq 5\%$ for sampling rate $\geq 50\%$ and almost exact recovery for sampling rate $\geq 60\%$, while MC and SRMF have error $10\% \sim 15\%$ at sampling rate $90\%$. Among all projects, the recovery errors of LS for IntelLab are lower than the other two, this is because the data matrices of InteLab have lower rank. More accurately, lower ratio of rank to matrix dimension, as shown in Fig. 5.

   Fig. 6 shows the recovery performance for the IntelLab project. For sampling rate $20\%$, $30\%$, $40\%$, and $50\%$, SRMF is worse than MC, since SRMF tries to fit the anomaly readings into a smooth surface. The recovery errors of MC and SRMF do not decrease further with more samples when sampling rate is larger than $60\%$ , LS and OS converge to zero recovery error since LS and OS take anomaly readings into consideration. Among the three physical conditions (humidity, light and temperature), the recovery error for humidity is the lowest for each case, because the humidity matrix has both lowest rank, as shown in Fig. 5, and lowest portion of anomaly readings, as shown in Fig. 4. Although the temperature matrix has relatively low rank too, but it also contains more anomaly readings.

   In Fig. 7 and 8, similar results hold for the GreenOrbs and NBDC CTD projects. Compared with IntelLab data matrices, we observe bigger gaps between MC/SRMF and LS/OS. The reasons are: (1) GreenOrbs and NBDC CTD have lower ratios of rank to matrix dimension, as shown in Fig. 5; (2) the dimensions, i.e., $min(N,T)$), of data matrices in GreenOrbs and NBDC CTD (i.e., $326$ and $216$, respectively), are much bigger than those in Intellab (i.e., $52$), as shown in Table I. For larger matrices, anomaly readings have weak influences. Note that the temperature of NBDC CTD has much lower recovery error than conductivity and salt, the reasons are: (1) the the rank of the temperature matrix is much lower, as shown in Fig. 5, and (2) the portion of anomaly readings is also lower.

\section{Conclusion}

  In this paper, we investigated the data recovery problem with considerations of massive missing entries, measurement noise, and anomaly readings. An LS-decomposition approach was proposed, which decomposed a given partially observed corrupted data matrix into a low-rank matrix and a sparse matrix. An accelerated proximal algorithm was devised to solve this problem. Theoretical results were given to guarantee the optimality and the converge rate. Our scheme is more robust than standard matrix completion methods, thus is expected to have better usability.
  %In the future, we will investigate how to apply the proposed LS-Decomposition scheme to event detection in wireless sensor networks.

\appendix

\section{Proofs}

\subsection{Proof of Theorem \ref{Main_results}}

 Our proof uses three crucial properties of $\langle \hat{L}, \hat{S} \rangle$:
 \begin{itemize}
 \item Since $\langle L,S\rangle$ is a feasible solution to (\ref{LSD}), we have:
 \begin{equation}
 ||\hat{L}||_{*} + \lambda ||\hat{S}||_{1} \leq ||L||_{*} + \lambda ||S||_{1}.
 \end{equation}
 \item The triangle inequality implies that:
 \begin{equation}\label{P_2delta}
 || \mathcal{P}_{\mathcal{O}} (\hat{L} + \hat{S} - L - S) ||_F
 \leq || \mathcal{P}_{\mathcal{O}} (\hat{L} + \hat{S} - R) ||_F  +
 || \mathcal{P}_{\mathcal{O}} (L + S - R) ||_F \leq 2\delta,
 \end{equation}
 since both $\langle \hat{L}, \hat{S} \rangle$ and $\langle L,S\rangle$ are feasible solutions to (\ref{LSD}).
 \item The condition we set in (\ref{condition}) to ensure a unique solution:
 \begin{equation}\label{0_condition}
  || \mathcal{P}_{\mathcal{O}^{\bot}(\hat{S})} ||_F =  || \mathcal{P}_{\mathcal{O}^{\bot}(S_0)} ||_F  = 0.
 \end{equation}
 \end{itemize}

 The first two properties imply that $\hat{X} = \langle \hat{L}, \hat{S} \rangle$ is close to $X = \langle L,S\rangle$. Set $\hat{X}=X + H$, where $H=\langle H_L, H_S \rangle$ and write $H^{\Gamma} = \mathcal{P}_{\Gamma}(H), H^{\Gamma^{\bot}} = \mathcal{P}_{\Gamma^{\bot}}(H)$ for short. Our aim is to bound $||H||_F$, which can be expressed as:
 \begin{equation}\label{HF_divide}
 ||H||_F^2 = ||H^{\Gamma}||_F^2 + ||H^{\Gamma^{\bot}}||_F^2
 = ||H^{\Gamma}||_F^2 +
 || (\mathcal{P}_{\mathcal{T}} \times \mathcal{P}_{\Omega}) (H^{\Gamma^{\bot}})||_F^2 + || (\mathcal{P}_{\mathcal{T}^{\bot}} \times \mathcal{P}_{\Omega^{\bot}} ) (H^{\Gamma^{\bot}})||_F^2.
 \end{equation}
 It suffices to bound each term in the righ-hand-side of (\ref{HF_divide}).

 \textbf{A. Bound the first term of (\ref{HF_divide})}.
 \begin{equation}\label{HF_T_divide}
 \begin{split}
 &||H^{\Gamma}||_F^2 = ||(H_L + H_S)/2||_F^2 + ||(H_L + H_S)/2||_F^2
 = \frac{1}{2}||H_L + H_S||_F^2 \\
 &= \frac{1}{2} \left( ||\mathcal{P}_{\mathcal{O}}(H_L + H_S)||_F^2 + ||\mathcal{P}_{\mathcal{O}^c}(H_L + H_S)||_F^2    \right) \\
 &=\frac{1}{2} \left( ||\mathcal{P}_{\mathcal{O}}(H_L + H_S)||_F^2 + ||\mathcal{P}_{\mathcal{O}^c}(H_L)||_F^2    \right)
 \end{split}
 \end{equation}
 where the last equation uses a condition drived from (\ref{0_condition}), i.e., $||\mathcal{P}_{\mathcal{O}^c}(H_S)||_F^2 =0$. For the first term of (\ref{HF_T_divide}), we already have $||\mathcal{P}_{\mathcal{O}}(H_L + H_S)||_F^2 \leq 4 \delta^2$ from (\ref{P_2delta}). Then, we bound the second term of (\ref{HF_T_divide}):
 \begin{equation}\label{HL_P_OC}
 ||\mathcal{P}_{\mathcal{O}^c}(H_L)||_F^2 = ||\mathcal{P}_{\mathcal{T}}(\mathcal{P}_{\mathcal{O}^c}(H_L))||_F^2 + ||\mathcal{P}_{\mathcal{T}^{\bot}}(\mathcal{P}_{\mathcal{O}^c}(H_L))||_F^2,
 \end{equation}
 and it suffices to bound each term in the right-hand-side.

 \textbf{We start with the second term of (\ref{HL_P_OC})}. Let $W$ be a dual certificate as in \cite{Wright2009,Wright2011,Wright2010Stable}. Then, $\Lambda = UV^{*} + W$ obeys $|| \mathcal{P}_{\mathcal{T}^{\bot}}(\Lambda)|| \leq 1/2$ and $||\mathcal{P}_{\Omega^{\bot}}(\Lambda)||_{\infty} \leq \lambda/2$. We have:
 \begin{equation}\label{L_HL}\small
  \setlength{\abovedisplayskip}{4pt}
 || L + H_L ||_{*} \geq || L + \mathcal{P}_{\mathcal{O}^c}(H_L) ||_{*} - || \mathcal{P}_{\mathcal{O}}(H_L)||_{*}
 \vspace{-4pt}
 \end{equation}
 and (\cite{candes2006near})
 \begin{equation}\label{L_P_HL}\small
  \setlength{\abovedisplayskip}{4pt}
 || L + \mathcal{P}_{\mathcal{O}^c}(H_L) ||_{*} \geq ||L||_{*} + (1 - ||\mathcal{P}_{\mathcal{T}^{\bot}}(\Lambda)||) ||\mathcal{P}_{\mathcal{T}^{\bot}}(\mathcal{P}_{\mathcal{O}^c}(H_L))||_{*}.
 \vspace{-4pt}
 \end{equation}
 Therefore, with $|| \mathcal{P}_{\mathcal{T}^{\bot}}(\Lambda)|| \leq 1/2$ and $||L+H_L||_{*} \leq ||L||_{*}$, combining (\ref{L_HL}) and (\ref{L_P_HL}) we have:
 \begin{equation}\small
  \setlength{\abovedisplayskip}{4pt}
 ||L||_{*} \geq || L||_{*} + \frac{1}{2} ||\mathcal{P}_{\mathcal{T}^{\bot}}(\mathcal{P}_{\mathcal{O}^c}(H_L))||_{*} - || \mathcal{P}_{\mathcal{O}}(H_L)||_{*}
 \vspace{-4pt}
 \end{equation}
 \begin{equation}\small
  \setlength{\abovedisplayskip}{4pt}
 ||\mathcal{P}_{\mathcal{T}^{\bot}}(\mathcal{P}_{\mathcal{O}^c}(H_L))||_{*} \leq 2||\mathcal{P}_{\mathcal{O}}(H_L)||_{*}.
 \vspace{-4pt}
 \end{equation}
 Since the nuclear norm dominated the Frobenius norm $||\mathcal{P}_{\mathcal{T}^{\bot}}(\mathcal{P}_{\mathcal{O}^c}(H_L))||_F \leq ||\mathcal{P}_{\mathcal{T}^{\bot}}(\mathcal{P}_{\mathcal{O}^c}(H_L))||_{*}$, we have
 \begin{equation}\small
  \setlength{\abovedisplayskip}{4pt}
 ||\mathcal{P}_{\mathcal{T}^{\bot}}(\mathcal{P}_{\mathcal{O}^c}(H_L))||_F \leq 2||\mathcal{P}_{\mathcal{O}}(H_L)||_{*} \leq 2\sqrt{n}||\mathcal{P}_{\mathcal{O}}(H_L)||_F
 \vspace{-4pt}
 \end{equation}
 where the second inequality follows from the Cauchy-Schwarz inequality. We know that
 \begin{equation}\small
  \setlength{\abovedisplayskip}{4pt}
 ||\mathcal{P}_{\mathcal{O}}(H_L)||_F \leq ||\mathcal{P}_{\mathcal{O}}(H_L + H_S)||_F \leq 2 \delta,
 \vspace{-4pt}
 \end{equation}
 therefore,
  \begin{equation}\label{HL_OC_TC}\small
   \setlength{\abovedisplayskip}{4pt}
 ||\mathcal{P}_{\mathcal{T}^{\bot}}(\mathcal{P}_{\mathcal{O}^c}(H_L))||_F \leq 4\sqrt{n}\delta.
 \vspace{-4pt}
 \end{equation}

  \textbf{We then bound $||\mathcal{P}_{\mathcal{T}}(\mathcal{P}_{\mathcal{O}^c}(H_L))||_F$ in the first term of (\ref{HL_P_OC})}. Observe that the assumption $\mathcal{P}_{\mathcal{T}} \mathcal{P}_{\mathcal{O}} \mathcal{P}_{\mathcal{T}} \succeq (p/2) \mathcal{I}$ together with $\mathcal{P}_{\mathcal{T}}^2 = \mathcal{P}_{\mathcal{T}},\mathcal{P}_{\mathcal{O}}^2 = \mathcal{P}_{\mathcal{O}} $, gives:
  \begin{equation}\label{Last_1}\small
  \begin{split}
   \setlength{\abovedisplayskip}{4pt}
  ||\mathcal{P}_{\mathcal{O}}\mathcal{P}_{\mathcal{T}}(\mathcal{P}_{\mathcal{O}^c}(H_L))||_F^2 &= \langle \mathcal{P}_{\mathcal{O}}\mathcal{P}_{\mathcal{T}}(\mathcal{P}_{\mathcal{O}^c}(H_L)),\mathcal{P}_{\mathcal{O}}\mathcal{P}_{\mathcal{T}}(\mathcal{P}_{\mathcal{O}^c}(H_L)) \rangle \\
  &=\langle \mathcal{P}_{\mathcal{O}}\mathcal{P}_{\mathcal{T}}(\mathcal{P}_{\mathcal{O}^c}(H_L)), \mathcal{P}_{\mathcal{T}}(\mathcal{P}_{\mathcal{O}^c}(H_L)) \rangle \\
  &\geq \frac{p}{2}||\mathcal{P}_{\mathcal{T}}(\mathcal{P}_{\mathcal{O}^c}(H_L))||_F^2.
  \vspace{-4pt}
  \end{split}
  \end{equation}
  But since
  \begin{equation}\small
   \setlength{\abovedisplayskip}{4pt}
  \mathcal{P}_{\mathcal{O}}(\mathcal{P}_{\mathcal{O}}) = 0= \mathcal{P}_{\mathcal{O}}\mathcal{P}_{\mathcal{T}}(\mathcal{P}_{\mathcal{O}^c}(H_L) + \mathcal{P}_{\mathcal{O}}\mathcal{P}_{\mathcal{T}^{\bot}}(\mathcal{P}_{\mathcal{O}^c}(H_L)),
  \vspace{-4pt}
  \end{equation}
  we have
  \begin{equation}\label{Last_2}\small
  \begin{split}
   \setlength{\abovedisplayskip}{4pt}
  ||\mathcal{P}_{\mathcal{O}}\mathcal{P}_{\mathcal{T}}(\mathcal{P}_{\mathcal{O}^c}(H_L))||_F &=||\mathcal{P}_{\mathcal{O}}\mathcal{P}_{\mathcal{T}^{\bot}}(\mathcal{P}_{\mathcal{O}^c}(H_L))||_F \\
  & \leq ||\mathcal{P}_{\mathcal{T}^{\bot}}(\mathcal{P}_{\mathcal{O}^c}(H_L))||_F.
  \vspace{-4pt}
  \end{split}
  \end{equation}
  Hence, (\ref{Last_1}) and (\ref{Last_2}) together give
  \begin{equation}\label{HL_T_OC}\small
   \setlength{\abovedisplayskip}{4pt}
 ||\mathcal{P}_{\mathcal{T}}(\mathcal{P}_{\mathcal{O}^c}(H_L))||_F^2 \leq \frac{2}{p}||\mathcal{P}_{\mathcal{T}^{\bot}}(\mathcal{P}_{\mathcal{O}^c}(H_L))||_F^2.
 \vspace{-4pt}
 \end{equation}
 Combining (\ref{HF_T_divide})(\ref{HL_P_OC})(\ref{HL_OC_TC})(\ref{HL_T_OC}) , we have:
 \begin{equation}\label{HT}
  \setlength{\abovedisplayskip}{4pt}
 ||H^{\Gamma}||_F^2 \leq 2\delta^2 + \frac{2+p}{p}8n\delta^2.
 \vspace{-4pt}
 \end{equation}

 \textbf{B. Bound the third term of (\ref{HF_divide})}.
 For any matrix pair $X=\langle L,S\rangle$, we define $||X||_{\diamondsuit}=||L||_{*} + \lambda ||S||_1$. We have 
 \begin{equation*}
 \begin{split}
 ||X + H||_{\diamondsuit} &\geq ||X + H^{\Gamma^{\bot}}||_{\diamondsuit} - ||H^{\Gamma}||_{\diamondsuit},\\
 ||X + H^{\Gamma^{\bot}}||_{\diamondsuit} &\geq ||X||_{\diamondsuit} + (3/4 - ||\mathcal{P}_{\mathcal{T}^{\bot}}(\Lambda)||) ||\mathcal{P}_{\mathcal{T}^{\bot}}(H_L^{\Gamma^{\bot}})||_{*} + (3\lambda/4 - ||\mathcal{P}_{\Omega^{\bot}}(\Lambda)||_{\infty}) ||\mathcal{P}_{\Omega^{\bot}}(H_S^{\Gamma^{\bot}})||_{*}\\
 & \geq ||X||_{\diamondsuit} + \frac{1}{4}(||\mathcal{P}_{\mathcal{T}^{\bot}}(H_L^{\Gamma^{\bot}})||_{*} + \lambda ||\mathcal{P}_{\Omega^{\bot}}(H_S^{\Gamma^{\bot}})||_{*} ),
 \end{split}
 \end{equation*}
 where the second inequality follows from Lemma 5 of \cite{Wright2010Stable}.
 
 Therefore, we have
 \begin{equation}
 ||\mathcal{P}_{\mathcal{T}^{\bot}}(H_L^{\Gamma^{\bot}})||_{*} + \lambda ||\mathcal{P}_{\Omega^{\bot}}(H_S^{\Gamma^{\bot}})||_{*} \leq 4 || H^{\Gamma} ||_{\diamondsuit}.
 \end{equation}
 For any matrix $Y \in \mathbb{R}^{n \times n}$, we have the following inequalities:
 \begin{equation}
 ||Y||_F \leq ||Y||_* \leq \sqrt{n}||Y||_F, ~~ \frac{1}{\sqrt{n}} ||Y||_F \leq \lambda ||Y||_1 \leq \sqrt{n} ||Y||_F,
 \end{equation}
 where we assume $\lambda = \frac{1}{\sqrt{n}}$. Then,
 \begin{equation}\label{Third_term}
 \begin{split}
 || (\mathcal{P}_{\mathcal{T}^{\bot}} \times \mathcal{P}_{\Omega^{\bot}} ) (H^{\Gamma^{\bot}})||_F^2 & \leq  || \mathcal{P}_{\mathcal{T}^{\bot}} (H_L^{\Gamma^{\bot}}) ||_F + || \mathcal{P}_{\Omega^{\bot}} (H_S^{\Gamma^{\bot}})  ||_F \\
 & \leq || \mathcal{P}_{\mathcal{T}^{\bot}} (H_L^{\Gamma^{\bot}}) ||_* + \lambda \sqrt{n}  || \mathcal{P}_{\Omega^{\bot}} (H_S^{\Gamma^{\bot}})  ||_1 \\
 & \leq 4 \sqrt{n} ||H^{\Gamma}||_{\diamondsuit} = 4 \sqrt{n} ( ||H_L^{\Gamma} ||_* + ||H_S^{\Gamma}||_1  )\\
 & \leq 4n ( ||H_L^{\Gamma} ||_F + ||H_S^{\Gamma}||_F ) = 4\sqrt{2}n ||H^{\Gamma}||_F
 \end{split}
 \end{equation}
 
 \textbf{C. Bound the second term of (\ref{HF_divide})}. By Lemma 6 of \cite{Wright2010Stable}, we have
 \begin{equation}\label{second_term_1}
 ||\mathcal{P}_{\Gamma} (\mathcal{P}_{\mathcal{T}} \times \mathcal{P}_{\Omega}) (H^{\Gamma^{\bot}})||_F^2 \geq \frac{1}{4}  || (\mathcal{P}_{\mathcal{T}} \times \mathcal{P}_{\Omega}) (H^{\Gamma^{\bot}})||_F^2.
 \end{equation}
 But since $\mathcal{P}_{\Gamma}(H^{\Gamma^{\bot}}) = 0 = \mathcal{P}_{\Gamma} (\mathcal{P}_{\mathcal{T}} \times \mathcal{P}_{\Omega}) (H^{\Gamma^{\bot}}) + \mathcal{P}_{\Gamma} (\mathcal{P}_{\mathcal{T}^{\bot}} \times \mathcal{P}_{\Omega^{\bot}}) (H^{\Gamma^{\bot}})$, we have
 \begin{equation}\label{second_term_2}
 \begin{split}
 ||\mathcal{P}_{\Gamma} (\mathcal{P}_{\mathcal{T}} \times \mathcal{P}_{\Omega}) (H^{\Gamma^{\bot}})||_F &= ||\mathcal{P}_{\Gamma} (\mathcal{P}_{\mathcal{T}^{\bot}} \times \mathcal{P}_{\Omega^{\bot}}) (H^{\Gamma^{\bot}})||_F \\
 & \leq ||(\mathcal{P}_{\mathcal{T}^{\bot}} \times \mathcal{P}_{\Omega^{\bot}}) (H^{\Gamma^{\bot}})||_F.
 \end{split}
 \end{equation}
 
 Combing (\ref{second_term_1}) and (\ref{second_term_1}), we have
 \begin{equation}
 || (\mathcal{P}_{\mathcal{T}} \times \mathcal{P}_{\Omega}) (H^{\Gamma^{\bot}})||_F^2 \leq 4|| (\mathcal{P}_{\mathcal{T}^{\bot}} \times \mathcal{P}_{\Omega^{\bot}} ) (H^{\Gamma^{\bot}})||_F^2,
 \end{equation}
 together with (\ref{HF_divide})(\ref{HT})(\ref{Third_term}), leads us to the final result:
 \begin{equation}\small
 \begin{split}
 ||H||_F^2 &\leq 5 \times 16 \times 2n^2 ||H^{\Gamma}||_F^2 + ||H^{\Gamma}||_F^2 \\
 &= 320n^2\delta^2(8n + 8n/p+ 1) + \frac{2+p}{p}8n\delta^2 + 2\delta^2.
 \end{split}
 \end{equation}

 Therefore, we obtain the desired result,
 \begin{equation}\small
  \setlength{\abovedisplayskip}{4pt}
 ||H||_F \leq (8n\sqrt{40n + 40n/p + 5}+ \sqrt{2})\delta.
 \vspace{-4pt}
 \end{equation}

\subsection{Proof of Theorem \ref{main_theorem}}

  First, we need the following three key inequalities.
  \begin{lemma}
  Let $\{a_k,b_k\}$ be positive sequences satisfying
  \begin{equation}\label{lemma:first}
  a_k - a_{k+1} \geq b_{k+1} - b_k, ~\forall k \leq 1, \text{with}~a_1 + b_1 \leq c,~c>0.
  %\vspace{-4pt}
  \end{equation}
  Then, $a_k \leq c$ for every $k \geq 1$.
  \end{lemma}

  By deduction, we can easily have:
  \begin{lemma}\label{lemma:second}
  The positive sequence $\{ t_k \}$ generated in our algorithm via $t_{k+1} \leftarrow \frac{1+\sqrt{4t_{k}^2+1}}{2}$ with $t_{-1} = 1$ satisfies $t_k \geq (k+1)/2$ for all $k \geq 1$.
  \end{lemma}

  Let $v_k \triangleq F(X_k) - F(X^{*}),~u_k \triangleq t_k X_k - (t_k - 1) X_{k_1} - X^{*}$. Similarly as is proved in \cite{Converge}, we have:
  \begin{lemma}
  The sequence $\{ X_k \}$ generated in our algorithm satisfies:
  \begin{equation}\label{lemma:third}
  t_k^2 v_k - t_{k+1}^2 v_{k+1} \geq ||u_{k+1} ||_F^2 - ||u_k||_F^2.
  %\vspace{-4pt}
  \end{equation}
  \end{lemma}

  Now, we are ready to prove our theorem. Let
  $a_k \triangleq t_k^2 v_k$, $b_k \triangleq ||u_k||_F^2$, $c \triangleq ||X_{k_0} - X^{*}||_F^2$,
  then from Lemma \ref{lemma:second},
  \begin{equation}
  a_k - a_{k+1} \geq b_{k+1} - b_k.
  %\vspace{-4pt}
  \end{equation}
  Assume that $a_{k} + b_{k} \leq c$ holds true for $k >1$,
  %$k \geq k_0 \triangleq \frac{log(\mu_0 / \mu)}{log(1/\eta)}$,
  combining Lemma \ref{lemma:first}, we obtain that:
  \begin{equation}
  t_k^2 v_k \leq ||X_{k} - X^{*}||_F^2,
  %\vspace{-4pt}
  \end{equation}
  which combined with Lemma \ref{lemma:second} yields
  \begin{equation}
  F(X_k) - F(X^{*}) \leq \frac{4||X_{1} - X^{*}||_F^2}{k^2}.
  %\vspace{-4pt}
  \end{equation}
  %where we treat $k_0, k_0+1,...,k$ as valid iterations.

\subsection{Proof of Lemma \ref{main_lemma}}

  From Theorem \ref{main_theorem}, we know that:
  \begin{equation}
  \lim_{k \rightarrow \infty} F(X_k) - F(X^{*}) \rightarrow 0,
  %\vspace{-4pt}
  \end{equation}
  which means that our algorithm converges to its globe optimal for large enough $k$.

  For any $\varepsilon > 0$, to guarantee $F(X_k)  - F(X^{*}) \leq \varepsilon$, then:
  \begin{equation}
  \frac{4||X_{1} - X^{*}||_F^2}{k^2} \leq \varepsilon,
  %\vspace{-4pt}
  \end{equation}
  \begin{equation}
  k \geq \frac{4||X_{1} - X^{*}||_F^2}{\sqrt{\varepsilon}} = O(1/\sqrt{\varepsilon}).
  %\vspace{-4pt}
  \end{equation}

\balance

% that's all folks
\end{document}